\documentstyle[12pt]{article} 
\input psfig.sty
\input epsfig.sty

\def\Title#1{\begin{center} {\Large #1 } \end{center}}
\def\Author#1{\begin{center}{ \sc #1} \end{center}}
\def\Address#1{\begin{center}{ \it #1} \end{center}}

\def\doeack{\footnote{Work supported by the Department of Energy,
                     contract DE--AC03--76SF00515.}}
\def\SLAC{Stanford Linear Accelerator Center\\
    Stanford University, Stanford, California 94309 USA}

\newenvironment{Abstract}{\begin{quotation} \begin{center}
                       ABSTRACT
     \end{center}\bigskip  }{\end{quotation}}
\def\beq{\begin{equation}}
\def\eeq#1{\label{#1}\end{equation}}
\def\eeqn{\end{equation}}
\def\beqa{\begin{eqnarray}}
\def\eeqa#1{\label{#1}\end{eqnarray}}
\def\eeqan{\end{eqnarray}}

\def\Re{{\cal R \mskip-4mu \lower.1ex \hbox{\it e}\,}}
\def\Im{{\cal I \mskip-5mu \lower.1ex \hbox{\it m}\,}}
\def\nn{\noindent}
\def\ie{{\it i.e.}}
\def\eg{{\it e.g.}}
\def\etc{{\it etc}}
\def\etal{{\it et al.}}

\def\sub#1{_{\lower.25ex\hbox{$\scriptstyle#1$}}}
\def\sul#1{_{\kern-.1em#1}}
\def\sll#1{_{\kern-.2em#1}}  
\def\sbl#1{_{\kern-.1em\lower.25ex\hbox{$\scriptstyle#1$}}}
\def\ssb#1{_{\lower.25ex\hbox{$\scriptscriptstyle#1$}}}
\def\sbb#1{_{\lower.4ex\hbox{$\scriptstyle#1$}}}

\def\to{\rightarrow}
\def\dk{\ifmmode \Delta\kappa\else $\Delta\kappa$\fi}
\def\sigt{\ifmmode \tilde\sigma\else $\tilde\sigma$\fi}
\def\mh{\ifmmode m\sbl H \else $m\sbl H$\fi}
\def\mch{\ifmmode m_{H^\pm} \else $m_{H^\pm}$\fi}
\def\mt{\ifmmode m_t\else $m_t$\fi}
\def\mc{\ifmmode m_c\else $m_c$\fi}
\def\mz{\ifmmode M_Z\else $M_Z$\fi}
\def\mw{\ifmmode M_W\else $M_W$\fi}
\def\mws{\ifmmode M_W^2 \else $M_W^2$\fi}
\def\mhs{\ifmmode m_H^2 \else $m_H^2$\fi}   
\def\mzs{\ifmmode M_Z^2 \else $M_Z^2$\fi}
\def\mts{\ifmmode m_t^2 \else $m_t^2$\fi}
\def\mcs{\ifmmode m_c^2 \else $m_c^2$\fi}
\def\mchs{\ifmmode m_{H^\pm}^2 \else $m_{H^\pm}^2$\fi}
\def\ztwo{\ifmmode Z_2\else $Z_2$\fi}
\def\zone{\ifmmode Z_1\else $Z_1$\fi}
\def\mtwo{\ifmmode M_2\else $M_2$\fi}
\def\mone{\ifmmode M_1\else $M_1$\fi}
\def\tb{\ifmmode \tan\beta \else $\tan\beta$\fi}
\def\xw{\ifmmode x\sub w\else $x\sub w$\fi}
\def\ch{\ifmmode H^\pm \else $H^\pm$\fi}
\def\lum{\ifmmode {\cal L}\else ${\cal L}$\fi}
\def\inpb{\ifmmode {\rm pb}^{-1}\else ${\rm pb}^{-1}$\fi}
\def\infb{\ifmmode {\rm fb}^{-1}\else ${\rm fb}^{-1}$\fi}
\def\epem{\ifmmode e^+e^-\else $e^+e^-$\fi}
\def\ppb{\ifmmode \bar pp\else $\bar pp$\fi}

\def\bsg{\ifmmode b\rightarrow s\gamma \else $b\rightarrow s\gamma$\fi}
 
\newskip\zatskip \zatskip=0pt plus0pt minus0pt
\def\matth{\mathsurround=0pt}

\def\atversim#1#2{\lower0.7ex\vbox{\baselineskip\zatskip\lineskip\zatskip
  \lineskiplimit 0pt\ialign{$\matth#1\hfil##\hfil$\crcr#2\crcr\sim\crcr}}}

\begin{document}
\rightline{\vbox{\halign{&#\hfil\cr
&SLAC-PUB-7365\cr
&December 1996\cr}}}
\vspace{0.8in} 
\Title{Extended Gauge Sectors at Future Colliders: Report of the New Gauge 
Boson Subgroup}
\bigskip
\Author{Thomas G. Rizzo\doeack}
\Address{\SLAC}
\bigskip

\begin{center}
{\bf Subgroup Members}: 
F. Cuypers (PSI),
S. Godfrey (Carleton), 
X-G. He (Melbourne), 
J. Hewett (SLAC), 
H. Kagan (OSU)$^\dagger$,
J. Lykken (FNAL), 
K. Maeshima (FNAL)$^\dagger$, 
L. Price (ANL), 
S. Riemann (Zeuthen), 
J. Rowe (U.C. Davis),
D. Toback (Chicago),
C-E. Wulz (Austria,OAW)
\end{center}
\bigskip

\begin{Abstract}
We summarize the results of the New Gauge Boson Subgroup on the physics of 
extended gauge sectors at future colliders as presented at the 1996 Snowmass 
workshop. We discuss the direct and indirect search reaches for new 
gauge bosons at both hadron and lepton colliders as well as the ability of 
such machines to extract detailed information on the couplings of these 
particles to the fermions and gauge bosons of the Standard Model. 
\end{Abstract}
\bigskip
\vskip1.0in
\begin{center}
To appear in the {\it Proceedings of the 1996 DPF/DPB Summer Study on New
 Directions for High Energy Physics-Snowmass96}, Snowmass, CO, 
25 June-12 July, 1996. 
\end{center}
%
\bigskip
\def\thefootnote{\fnsymbol{footnote}}
\setcounter{footnote}{0}
\newpage
\section{Introduction}
\subsection{Overview}

One of the most important goals of existing and future colliders is to 
establish the gauge group which fully describes the strong and electroweak 
interactions. Current precision measurements{\cite {warsaw}} as well as 
direct collider searches{\cite {tev}}, both of which probe the physics at 
the `electroweak' (100 GeV) 
scale, support the hypothesis that this group is that of the Standard 
Model(SM): $SU(3)_c\times SU(2)_L \times U(1)_Y$. Many scenarios have been 
proposed over the last 25 years in which the SM is just an effective low 
energy version of a somewhat more complex gauge structure which exists at 
higher energies. If any of these ideas have any validity and the associated 
scale is not far above the multi-TeV range then future colliders should find 
direct evidence for its existence. 
There are many reasons why the discovery of such a new scale would be 
important. Perhaps the most obvious is the observation that we cannot 
extrapolate the physics we currently see to extremely high energies, such as 
the Planck or a hypothetical 
GUT scale, without knowing all that is happening in our own neighborhood that 
we are just beginning to probe. Clearly, 
the discovery of a new gauge boson, such as a $Z'$ or $W'$, would be the 
cleanest signature for new physics beyond the SM.

It is impossible in a brief review to cover all possible models with new gauge 
bosons. We note that extensions of both the strong and electroweak sectors have 
been proposed in a variety of forms. For example, extending the conventional 
QCD $SU(3)_c$ group to $SU(3)_1\times SU(3)_2$ leads to scenarios which 
predicts new strongly interacting particles such as axigluons{\cite {axi}}, 
colorons{\cite {col}}, and topgluons{\cite {tc}} depending upon how the 
quarks transform under the two $SU(3)$'s. Other possibilities include 
extending the color group to larger factors, such as $SU(5)_c${\cite {foot}}. 
{\it All} of these extensions result in particles which are new 
gauge bosons in the strictest sense. As the physics of such states are 
covered in New Interactions 
Subgroup report{\cite {harris}}, we will limit our discussion below to 
extensions of the SM electroweak group. Even with this constraint, the number 
of potential models remains very large.

Extended Gauge Models(EGMs) can be divided into two very broad classes 
depending upon whether or not they originate from a GUT group, such 
as $SO(10)$ or 
$E_6$. {\it Generally}, the new gauge bosons from GUT-inspired scenarios have 
generation-independent couplings (in the same sense as the $W$ and $Z$ of the 
SM), whereas this need not be true for non-unifiable models. Also, 
{\it generally}, the extension of the SM group structure induces additional 
anomalies which cannot be cancelled by using the conventional SM fermions 
alone. This implies the almost all EGMs also contain additional exotic matter 
particles, such as leptoquarks, with masses comparable to those of the new 
gauge bosons themselves. In what follows, we will limit our discussion 
almost exclusively to a small set of sample models of either class that have 
been recently reviewed by Cvetic and Godfrey{\cite {rev}}. 

The search reach at a collider as well as our ability to extract coupling 
information for a new gauge boson is somewhat model 
dependent due to the rather large variations in their 
couplings to the SM fermions. To be specific we 
consider ({\it i}) the $E_6$ effective rank-5 model(ER5M), which predicts a 
$Z'$ whose couplings depend on a single parameter 
$-\pi/2 \leq \theta \leq \pi/2$, with models $\psi (\theta=0)$, 
$\chi (\theta=-\pi/2)$, $I(\theta=-\cos^{-1}\sqrt {3/8})$, and 
$\eta (\theta= \cos^{-1}\sqrt {5/8})$ 
denoting specific common cases discussed in the literature; 
({\it ii}) the Sequential Standard Model(SSM) 
wherein the new $W'$ and $Z'$ are just heavy versions of the SM particles (of 
course, this is not a true model in the strict sense but is commonly used as a 
guide by experimenters); ({\it iii}) the Un-unified Model(UUM), based on the 
group $SU(2)_\ell \times SU(2)_q \times U(1)_Y$, which has a 
single free parameter $0.24 \leq s_\phi \leq 0.99$; 
({\it iv}) the Left-Right Symmetric Model(LRM), based on the group 
$SU(2)_L \times SU(2)_R \times U(1)_{B-L}$, 
which also has a free parameter ($\kappa=g_R/g_L\geq 0.55$) of order unity 
which is just the ratio of the gauge couplings 
and, lastly, ({\it v}) the Alternative Left-Right Model(ALRM), based on the 
same extended group as the LRM but now arising from 
$E_6$, wherein the fermion assignments are modified in comparison to the LRM 
due to an ambiguity in how they are embedded in the {\bf 27} representation. 

In the case of a $W'$ we will restrict ourselves to the specific example of 
the LRM, \ie, $W_R$, although both the UUM and ALRM have interesting $W'$ 
bosons. 
The $W'$ in the UUM is quite similar to that of the SSM apart from its overall 
coupling strength and the size of its leptonic branching fraction. The $W'$ 
in the ALRM cannot be singly produced via the Drell-Yan mechanism since it 
carries non-zero lepton number and negative $R-$parity{\cite {physrep}}. 
In what follows $Z-Z'$ and $W-W'$ mixing effects will be generally 
ignored which 
is an excellent approximation for any new gauge bosons in the multi-TeV mass 
range.

\subsection{Why a $Z'$ Might Be Light}

While it is interesting to consider EGMs on their own merits, they are only of 
true phenomenological interest if their associated scale is within the range 
accessible to existing or future colliders. In principle, the new scale could 
lie anywhere between the electroweak scale and the Planck scale. If it is far 
from current energies then the associated new physics could only be observed 
indirectly. Why might we expect this scale to be `nearby'? In a contribution 
to these proceedings, Lykken{\cite {lykken}} examined this issue for the case 
of a new $U(1)'$ gauge group within the general context of SUSY-GUTS and 
String Theory with weak-scale supersymmetry, extending the work of Cvetic and 
Langacker{\cite {cl}}. 

One of the essential ingredients of this scenario is the idea of radiative 
symmetry breaking. It is easy to imagine that the breaking of the $U(1)'$ 
might be triggered by the renormalization group(RG) running of some exotic 
fermion fields which drive the mass squared of some exotic scalar field 
negative. {\it However}, due to the logarithmic nature of the RG running it 
would not seem very likely that the $Z'$ mass would naturally lie in the 
few TeV region or below without some fine tuning of parameters. In fact, in 
scenarios of this kind, one finds that the $Z'$ mass naturally lies instead 
in the $10^8-10^{16}$ GeV range for typical GUT models. 

In the MSSM, symmetry breaking is induced by the vev's of the two Higgs 
doublets $H_{U,D}$. To break $U(1)'$, we require the introduction of some 
number of SM singlet fields of which at least one gets a vev. In models with 
two or more singlets getting vev's, D flatness imposes a relationship between 
these vev's (apart from corrections of order the soft SUSY breaking scale) 
but does not relate them to the vev's of $H_{U,D}$. This implies 
that the $Z'$ mass and the electroweak scale are not directly related and the 
$Z'$ could naturally be quite massive. On the otherhand, if only one 
singlet ($S$) gets 
a vev and either or both of $H_{U,D}$ carry $U(1)'$ charges then the doublet 
and singlet vev's are directly related through the requirement of D flatness. 
If the soft mass for $S$, $m_S^2$, is of order the weak scale (as is the case 
for all SUSY breaking soft terms) then the vev of $S$ is also of order the 
electroweak scale. The $Z'$ mass then becomes calculable in terms of the 
vev's, which are no longer independent, the gauge couplings, and the $U(1)'$ 
charges of the singlet and doublet fields. 

To this scenario certain phenomenological constraints need to be added in 
that ($i$) the $Z'$ has to be sufficiently massive as to have avoided 
current searches and ($ii$) either the $Z-Z'$ mixing 
angle must be reasonably small, of order $10^{-3}$, or the $Z'$ couplings 
to leptons are suppressed (\ie, the $Z'$ is leptophobic). This second 
constraint arises from the excellent agreement between leptonic precision 
measurements at SLD and LEP and the predictions of the SM. If the $Z'$ is 
not leptophobic, 
this constraint implies an additional strong constraint between the $U(1)'$ 
charges of $S$, $Q'_S$, and the Higgs doublets, $Q'_H$. Cvetic and 
Langacker{\cite {cl}} did 
not find an acceptable string model of this type amongst those presented in 
the literature; of course only a handful of such models are known so far. (We 
recall that within these string models all of the $U(1)'$ charges are 
completely specified.) Given the severity of the constraint this is probably 
not surprising. Lykken examined the possibility of a leptophobic $Z'$ in 
the string context with the additional requirement that $Q'_H$ be non-zero. 
His results are shown in Table~\ref{lyk}. As can be seen, only Flipped $SU(5)$ 
is a potential candidate theory but a detailed study{\cite {lykken}} shows 
that the particle embedding necessary to generate a large top Yukawa coupling 
lead to flavor changing neutral currents generated by $Z'$ exchange. Lykken 
concludes that a leptophobic $Z'$ satisfying our constraints is less natural 
in string theory than the more conventional kind of $Z'$.
Finally, Lyyken further reminds us that the $U(1)'$ leads to potentially 
large D-term contributions to the squark and slepton masses which can be of 
order 250 GeV. If so this implies rather significant modifications in the 
sparticle mass spectrum in comparison to either minimal supergravity or 
gauge-mediated low energy breaking models.

\begin{table}
\begin{center}
\begin{tabular}{c|cc}
\hline
Model & Leptophobic $U(1)'$? & $Q'_H$ $\neq$ 0? \cr
\hline
Faraggi I\ \cite{farI}&no&-- \cr
Faraggi II\ \cite{farII}&yes&no \cr
Faraggi et al\ \cite{fny}&no&-- \cr
Chaudhuri et al\ \cite{chl}&yes&no \cr
Hockney-Lykken\ \cite{hl}&yes&no \cr
Flipped SU(5)\ \cite{flip}&yes&yes\cr
\hline
\end{tabular}
\label{lyk}
\caption{Partial survey of string models for naturally light $Z'$ candidates 
which are leptophobic from Ref.10.}
\end{center}
\end{table}

\section{Collider Search Reaches For New Gauge Bosons}

\begin{figure}[htbp]
\centerline{
\psfig{figure=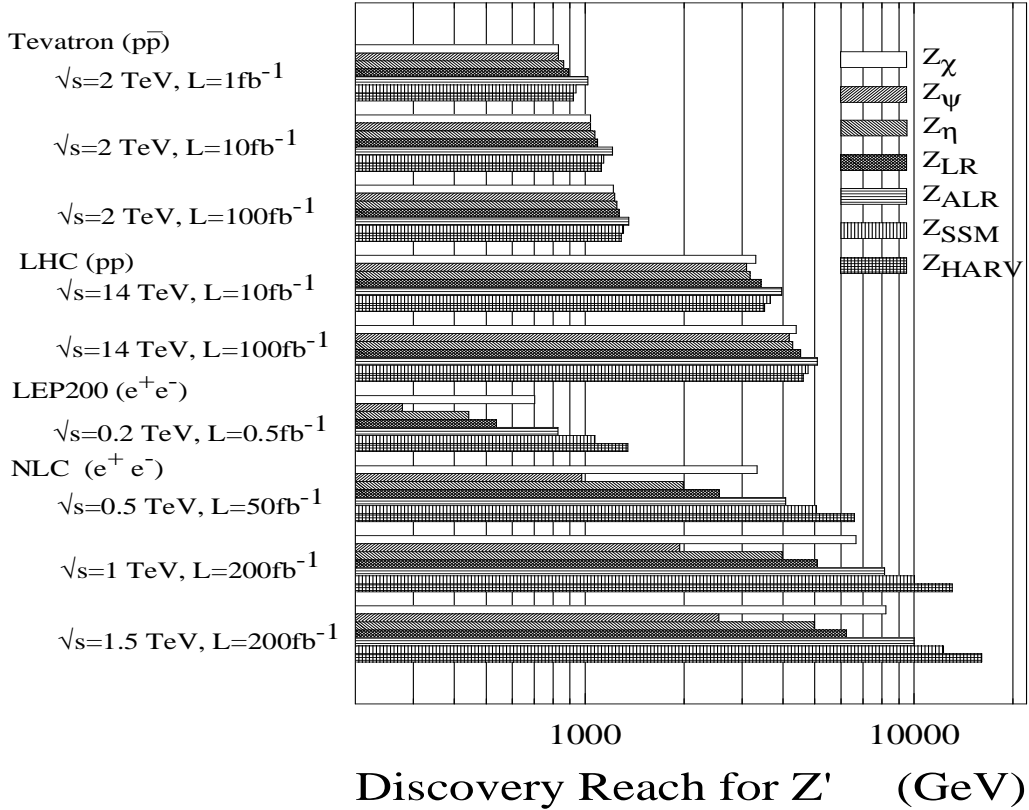,height=15cm,width=14cm,angle=0}}
\vspace*{-1.9cm}
\caption{ Tevatron and LHC bounds are based on 10 events in the 
$e^+e^-+\mu^+\mu^-$ channels; decays to SM final states only is assumed. LEP 
and NLC bounds are $99\%$ CL using the observables 
$\sigma_l,~R^{had},~A_{LR}^l$ and $A_{LR}^{had}$.}
\label{steve}
\end{figure}

The search capabilities for new gauge bosons of existing and future 
accelerators as been discussed by many authors and has been most recently 
summarized by Fig.~\ref{steve} from Cvetic and Godfrey{\cite {rev}}. More 
recent work along these lines was presented as this meeting{\cite {tgr}} 
which generalize and extend previous results. The discussion for hadron and 
lepton colliders are presented in subsequent sections. 

\vspace*{-0.5cm}
\nn
\vspace*{0.4mm}

\subsection{Hadron Colliders}

In what follows we will mostly limit our analysis to the conventional 
discovery channels involving $Z'$ and $W'$ decays to charged lepton pairs and 
charged leptons plus missing $E_t$, respectively. 
Regrettably, this leaves many territories untouched 
wherein, \eg, the new gauge boson decays to dijets, pairs of SM gauge bosons, 
or leptonic $W'$ decay modes not involving missing $E_t$. 
(Toback{\cite {toback}} has 
partially remedied this situation for the Tevatron as will be discussed 
below.) These possibilities require further study particularly at the LHC.

\vspace*{-0.5cm}
\nn
\begin{figure}[htbp]
\centerline{
\psfig{figure=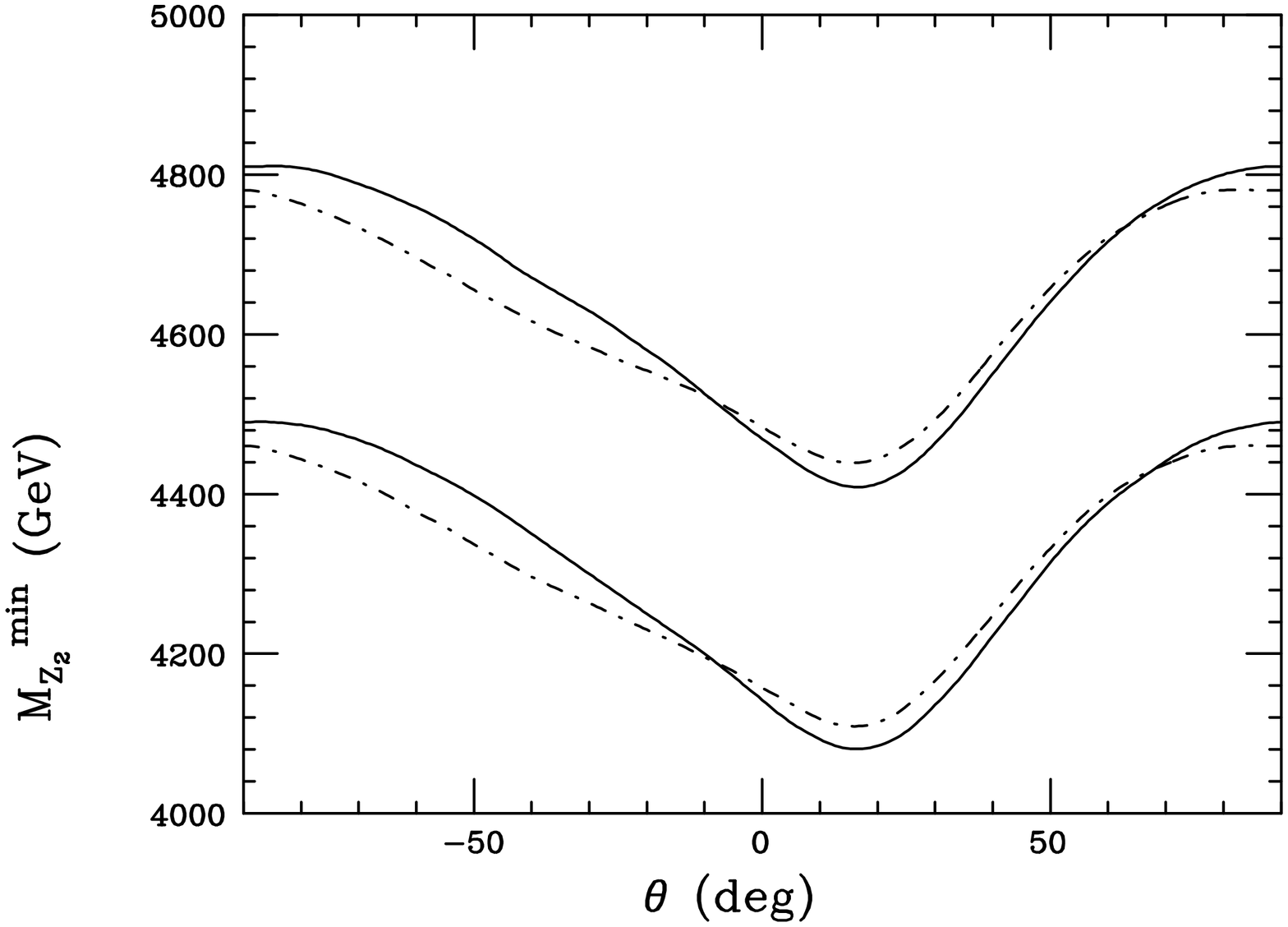,height=9.1cm,width=9.1cm,angle=-90}
\hspace*{-5mm}
\psfig{figure=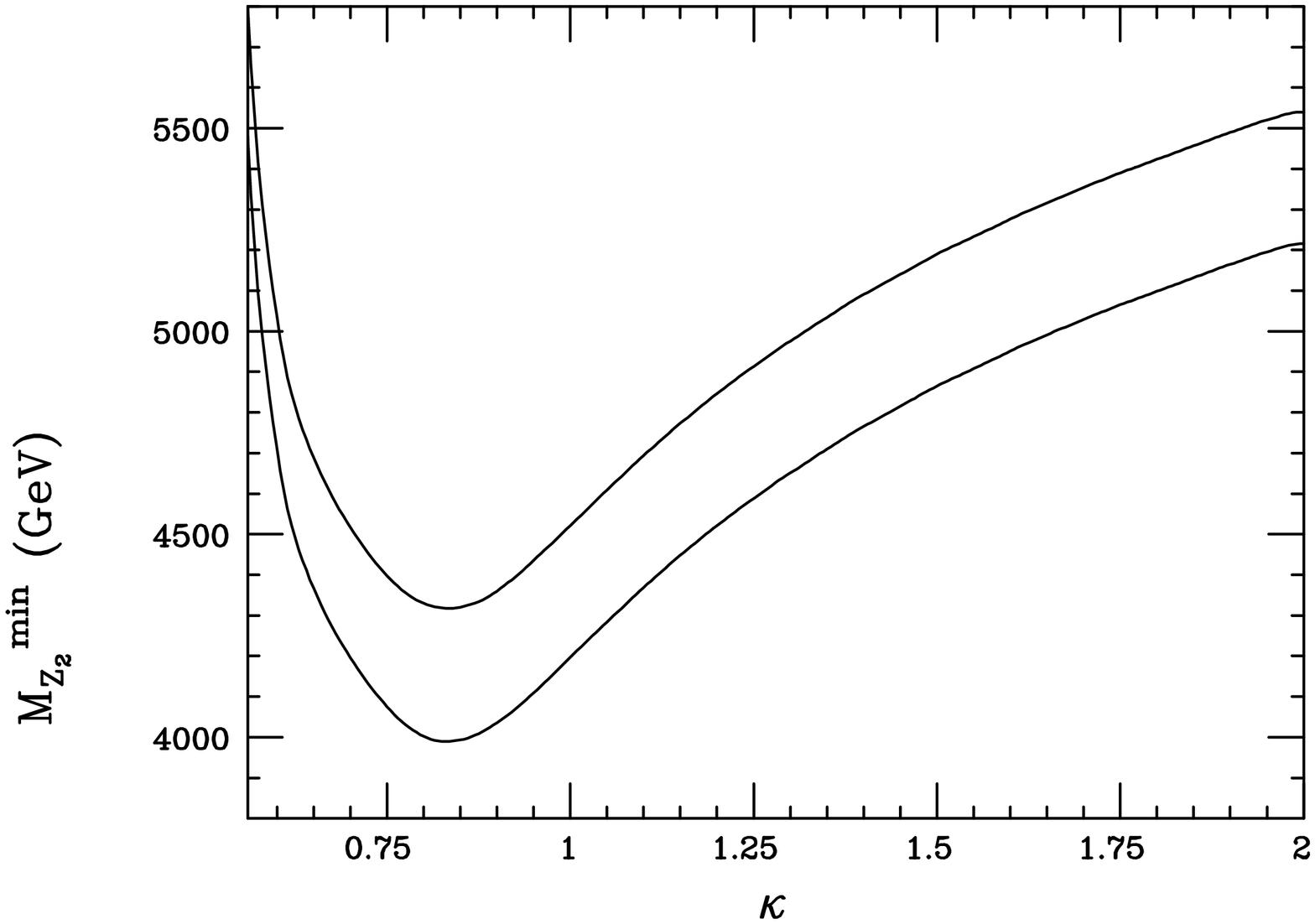,height=9.1cm,width=9.1cm,angle=-90}}
\vspace*{-0.6cm}
\caption{$Z'$ search reaches at the 14 TeV LHC for $E_6$ 
models as a function of $\theta$(left) and for the LRM as a function of 
$\kappa$(right). The curves on the left(right) correspond to integrated 
luminosities 100 and 200(50 and 100) $fb^{-1}$, respectively. }
\label{figlhc}
\end{figure}
\vspace*{0.1mm}

Traditionally, both $Z'$ and $W'$ search reaches are obtained through the 
use of the narrow width 
approximation with some additional corrections to account for detector 
acceptance's($A$) and efficiencies($\epsilon$). In this case the number of 
expected events($N$) is simply the product 
$N=\sigma B_l A \epsilon {\cal L}$, where $\sigma$ is the 
production cross section, $B_l$ is the leptonic branching fraction and 
${\cal L}$ is the machine's integrated luminosity. 
A $5\sigma$ signal is assumed to be given by 10 signal events with no 
background; this is logically consistent since an extremely narrow peak in the 
dilepton mass distribution can have only an infinitesimal background 
underneath it. Detailed detector simulations for both the Tevatron and 
LHC{\cite {wulz}} validate this 
approximation as a good estimator of the true search reach at least for the 
more `traditional' models where the $Z'$ and $W'$ are relatively narrow. 

\vspace*{-0.5cm}
\nn
\begin{figure}[htbp]
\centerline{
\psfig{figure=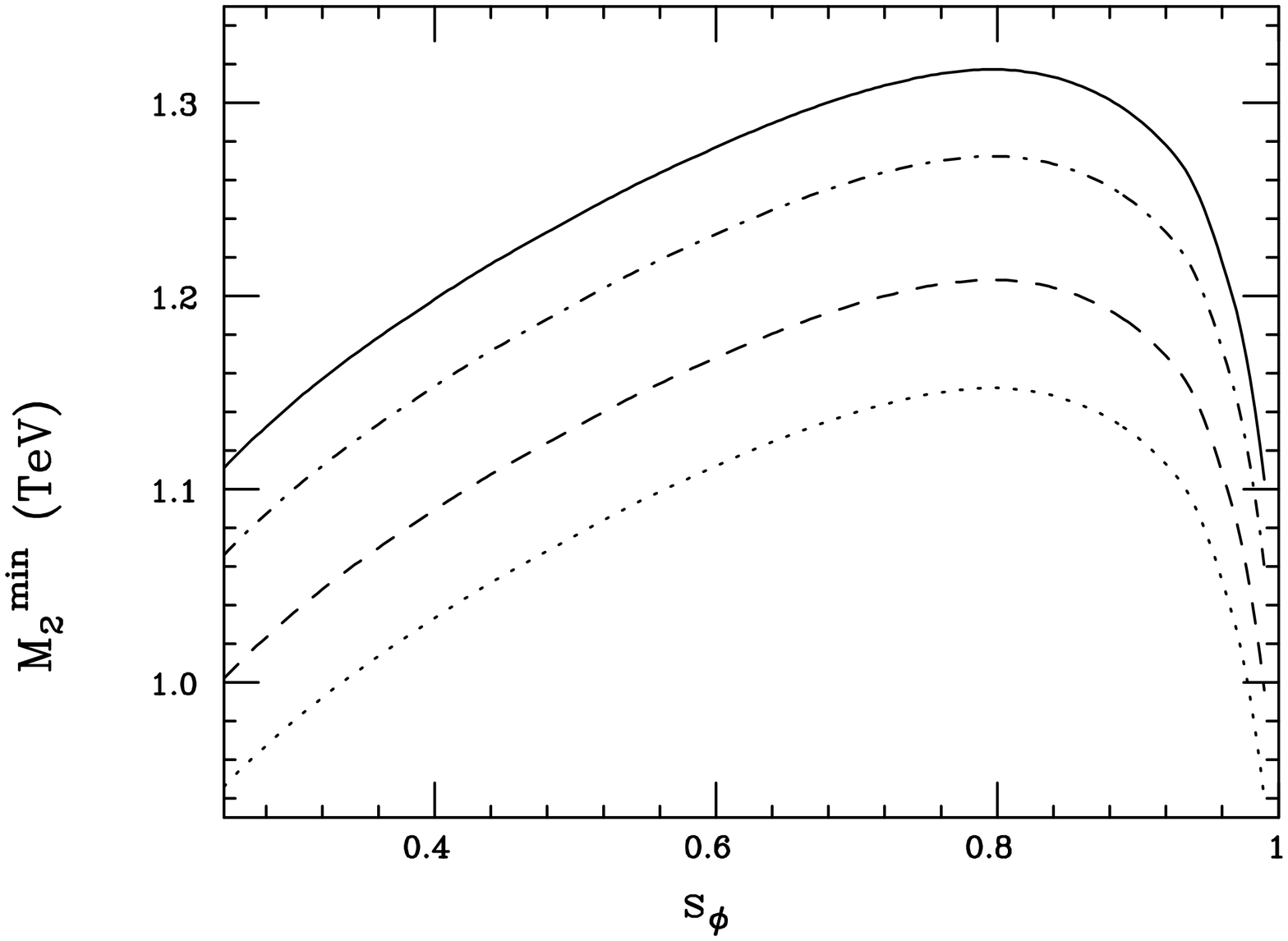,height=9.1cm,width=9.1cm,angle=-90}
\hspace*{-5mm}
\psfig{figure=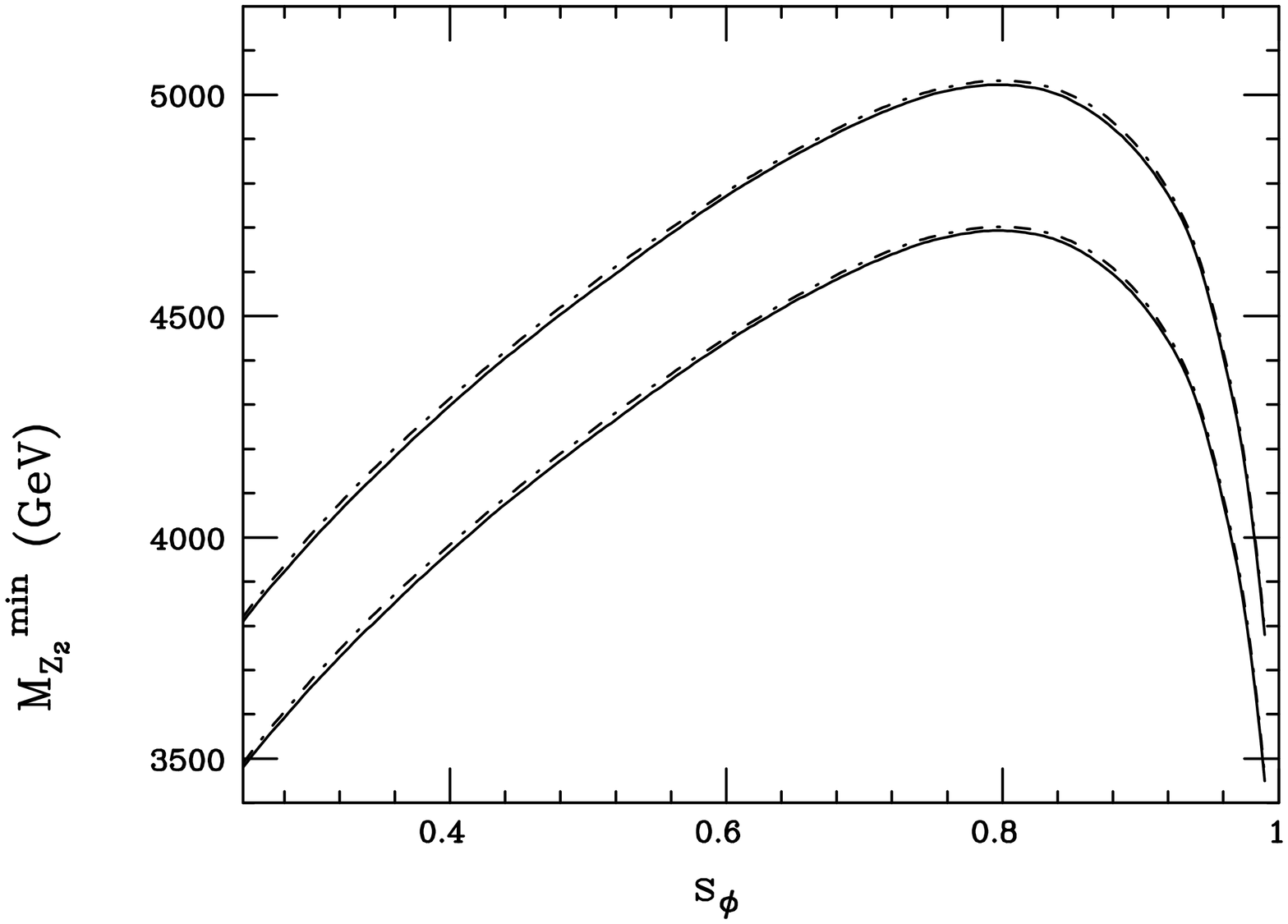,height=9.1cm,width=9.1cm,angle=-90}}
\vspace*{-0.6cm}
\caption{ Same as the previous figure, but now for the UUM. On the left are 
the results for the Tevatron running at 
2 TeV. From top to bottom the integrated luminosities are assumed to be 100, 
50, 20 and 10 $fb^{-1}$, respectively. On the right are the corresponding LHC 
results for 50 and 100 $fb^{-1}$.}
\label{figuum}
\end{figure}
\vspace*{0.1mm}

To obtain the search reach in the $Z'$ case, 
we need to know the various fermionic couplings for a fixed value of the 
$Z'$ mass to obtain $\sigma$. Traditionally, one also assumes that the 
$Z'$ can {\it only} decay to pairs of SM fermions in order to obtain 
$B_l$. It is important to note that in many models, 
where the $Z'$ can also decay to exotic fermions and/or 
SUSY particles this {\it overestimates} $B_l$ and, thus, the search reach. In 
obtaining our results for 10 signal events we combine both the electron and 
muon decay channels. 
With these assumptions, Figure~\ref{figlhc} shows the 
discovery reaches for the $Z'$ of the ER5M and the LRM at the LHC whereas 
Figure~\ref{figuum} shows the corresponding reaches for the UUM $Z'$ at both 
the Tev33 and the LHC. The full set of figures for other models/colliders 
can be found in  Ref.{\cite {tgr,bsm}}. 
Table~\ref{$Z'$ lhc} contains a summary of all of these results. Here we 
see that TeV33 will allow us to approach the 1 TeV mass scale for $Z'$ bosons 
for the first time. Note that 
in the case of the 60 and 200 TeV machines the higher $q\bar q$ 
luminosities in the $p\bar p$ mode leads to a significantly 
greater ($\simeq 30-50\%$) search reach.

\begin{table*}[htpb]
\leavevmode
\begin{center}
\label{$Z'$ lhc}
\begin{tabular}{lcccccc}
\hline
\hline
Model & LHC&60 TeV ($pp$)&60 TeV ($p\bar p$)&200 TeV ($pp$)&
200 TeV ($p\bar p$)&TeV33 \\
\hline
$\chi$  & 4.49 & 13.3 & 17.5 & 43.6 & 63.7 & 1.00    \\
$\psi$  & 4.14 & 12.0 & 17.1 & 39.2 & 62.3 & 1.01    \\
$\eta$  & 4.20 & 12.3 & 17.9 & 40.1 & 64.8 & 1.03    \\
I       & 4.41 & 12.9 & 15.2 & 42.1 & 56.0 & 0.88    \\
SSM     & 4.88 & 14.4 & 20.6 & 45.9 & 68.7 & 1.10     \\
ALRM    & 5.21 & 15.0 & 22.5 & 49.9 & 74.7 & 1.15     \\
LRM     & 4.52 & 13.5 & 18.9 & 43.2 & 64.6 & 1.05     \\
UUM     & 4.55 & 13.7 & 19.7 & 43.5 & 65.1 & 1.08     \\
        & & & & & & \\
Hit     & 0.33 & 1.5 & 1.8 & 4.9 & 6.3 & 0.05 \\
\hline
\hline
\end{tabular}
\caption{$Z'$ search reaches at hadron colliders in TeV. For the LRM, 
$\kappa=1$ is assumed while for the UUM, we take $s_\phi=0.5$. Decays to only 
SM fermions is assumed. The luminosities of the Tevatron, LHC, 60 TeV and 
200 TeV colliders are assumed to be 10, 100, 100 and 1000 $fb^{-1}$, 
respectively. The last line in the Table is the approximate reduction in 
reach in TeV due to a decrease in $B_l$ by a factor of 2.}
\end{center}
\end{table*}

If the above estimate of the leptonic branching fraction is wrong, how 
seriously are the search reaches compromised? To get a feeling for this, 
consider reducing the 
value of $B_l$ by a factor of two from the naive estimate given by 
the assumption that the $Z'$ decays to 
only SM fermion pairs. (In the $E_6$ case, this roughly corresponds to 
allowing the $Z'$ to decay into SUSY partners as well as the exotic fermions 
with some phase space suppression{\cite {physrep}}.) 
Semi-quantitatively, the reduction in reach for each collider is found to be 
roughly model independent and approximate results are given in the last line of 
Table~\ref{$Z'$ lhc}. As can be seen from these values the `hit' taken can be 
significant in some cases. However, unless $B_l$ is very much smaller than the 
naive estimate it is clear that the multi-TeV mass range will remain 
easily accessible to future hadron colliders.

\vspace*{-0.5cm}
\nn
\begin{figure}[htbp]
\centerline{
\psfig{figure=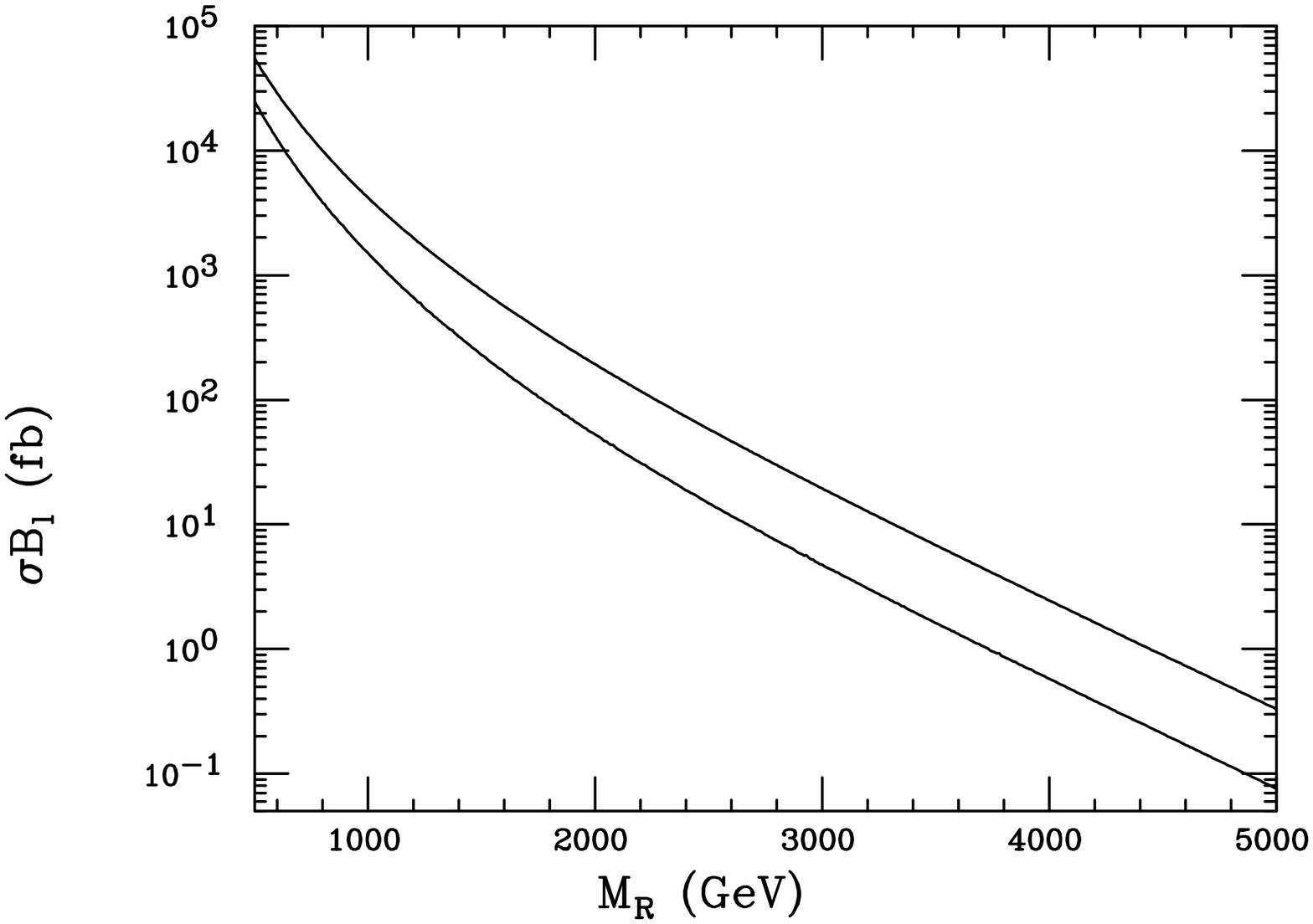,height=9.1cm,width=9.1cm,angle=-90}
\hspace*{-5mm}
\psfig{figure=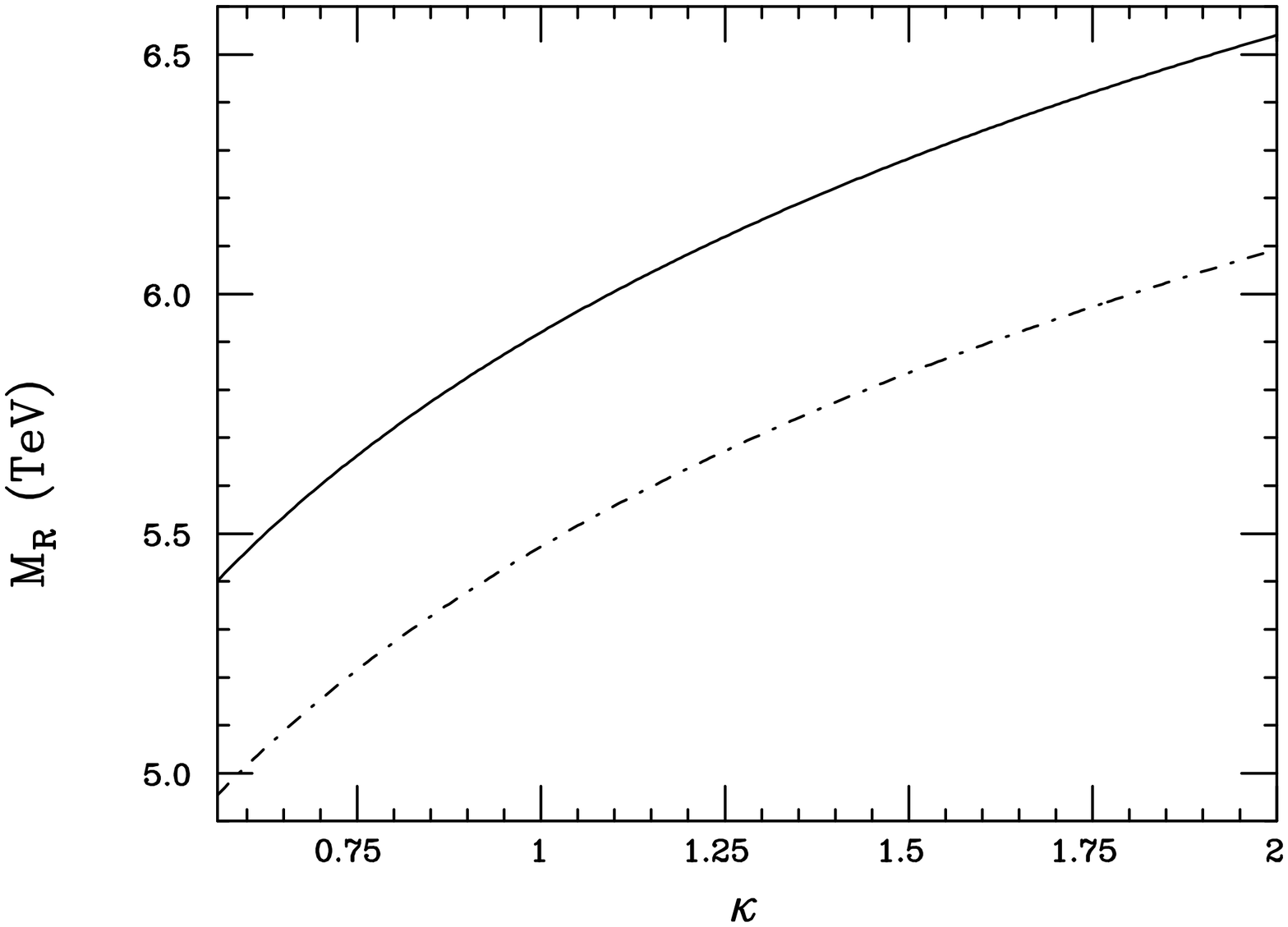,height=9.1cm,width=9.1cm,angle=-90}}
\vspace*{-0.6cm}
\caption{$W_R$ production cross section times leptonic branching fraction 
at the LHC(left) for $\kappa=1$ assuming $V_L=V_R$(top) or the worst case 
values(lower) of $V_R$. Also shown is the search reach for $W_R$ vs. 
$\kappa$(right) at the 
LHC with $V_L=V_R$ for luminosities of 50 and 100 $fb^{-1}$.}
\label{wr}
\end{figure}
\vspace*{0.1mm}

Unlike the $Z'$ case, the corresponding $W_R$ searches in the LRM 
via the Drell-Yan process have many subtleties even when we assume that the 
missing $E_t$ mode is accessible and dominant. The canonical search assumes 
that the $q'\bar qW_R$ 
production vertex has SM strength, implying ({\it i}) $\kappa=1$ and 
({\it ii}) $|V_{L_{ij}}|=|V_{R_{ij}}|$, \ie, the elements of the RH CKM mixing 
matrix, $V_R$, are the same as $V_L$, and, as in the $Z'$ case, ({\it iii})  
that the $W_R$ leptonic branching fraction is given by its decay to SM 
fermions only. Of course violations of 
assumptions ({\it i}) and ({\it iii}) are easily accounted for in a manner 
similar to the $Z'$ case discussed above. 
If assumption ({\it ii}) is invalid, a significant search reach degradation 
can easily occur as a result 
of modifying the weight of the various parton luminosities which enter into 
the calculation of the production cross section. At the $pp$ colliders such 
as the LHC, we do not expect that surrendering ({\it ii}) 
will cost us such a very large penalty since the $W_R$ production process 
already occurs 
through the annihilation of sea$\times$valence quarks. On the otherhand, $W_R$  
production is a valence$\times$valence process at the $p\bar p$ colliders 
such as the Tevatron so we might anticipate a more significant reach 
reduction in this case. If the conventional $W'$ decay modes are suppressed, it
may also be wise to search for the $WZ$ final state as discussed 
by {\cite {toback}}.  

In models where both a $W'$ and a $Z'$ exist there is generally a direct 
relationship between their masses. For example, in the UUM case the $W'$ and 
$Z'$ are predicted to be degenerate, whereas in the LRM there is a non-trivial 
relationship:
\begin{equation}
{M_{Z_R}^2\over {M_{W_R}^2}}={\kappa^2(1-x_w)\rho_R\over {\kappa^2(1-x_w)
-x_w}} \,,
\end{equation}
where $\rho_R=1(2)$ signal symmetry breaking of $SU(2)_R$ by right-handed Higgs 
doublets(triplets) and $x_w=\sin^2 \theta_w$. A measurement of the $W'$ to 
$Z'$ mass ratio will tell us a fair amount about the underlying gauge theory 
extension.

%
\begin{table}[htpb]
\begin{center}
\label{$W_R$ lhc}
\begin{tabular}{lcc}
\hline
\hline
Machine  & $V_L=V_R$  & $V_R$ (WC) \\
\hline
TeV33               &  1.2 & $\simeq 0.5$\\
LHC                 &  5.9 &  5.1 \\
60 TeV ($pp$)       & 19.7 & $\simeq 16$ \\
60 TeV ($p\bar p$)  & 25.1 & $\simeq 16$ \\
200 TeV ($pp$)      & 64.7 & $\simeq 52$ \\
200 TeV ($p\bar p$) & 82.9 & $\simeq 52$ \\
\hline
\hline
\end{tabular}
\caption{$W_R$ search reaches of hadron colliders in the lepton plus 
missing energy mode in TeV. 
$\kappa=1$ and decays to only SM fermions is assumed. WC(worst case) 
refers to the set of 
$V_R$ elements that yield the lowest production cross section. The 
luminosities are as in the previous Table.}
\end{center}
\end{table}

Fig.~\ref{wr} summarizes the $W_R$ search at the LHC where the narrow width 
approximation has been employed. In particular this figure 
shows that the reduction of reach at the LHC due to variations in $V_R$ is 
rather modest whereas it is far more significant at the Tevatron. 
The corresponding figures for the complete set of results at other colliders 
can be found in  Ref.{\cite {tgr,bsm}}; Table~\ref{$W_R$ lhc} summarizes 
these findings. We note that for the case of $W_R$, if we let 
$B_\ell \to B_\ell/2$, the search reach at the LHC is reduced by $\simeq 450$ 
GeV for values of $\kappa$ in the range $0.55 \leq \kappa \leq 2$.

\subsection{Lepton Colliders: Indirect Searches}

It is more than likely that a $Z'$ will be too massive to be produced directly 
at the first generation of new lepton colliders. Thus searches at such 
machines will be indirect and will consist of 
looking for deviations in the predictions of the SM in as many observables as 
possible. Layssac \etal {\cite {rev}} have shown that the deviations in the 
leptonic observables due to the existence of a $Z'$ are rather unique. Since 
the $Z'$ is not directly produced, lepton collider searches are insensitive to 
the decay mode assumptions that we had to make in the case of hadron colliders.

\vspace*{-0.5cm}
\nn
\begin{figure}[htbp]
\centerline{
\psfig{figure=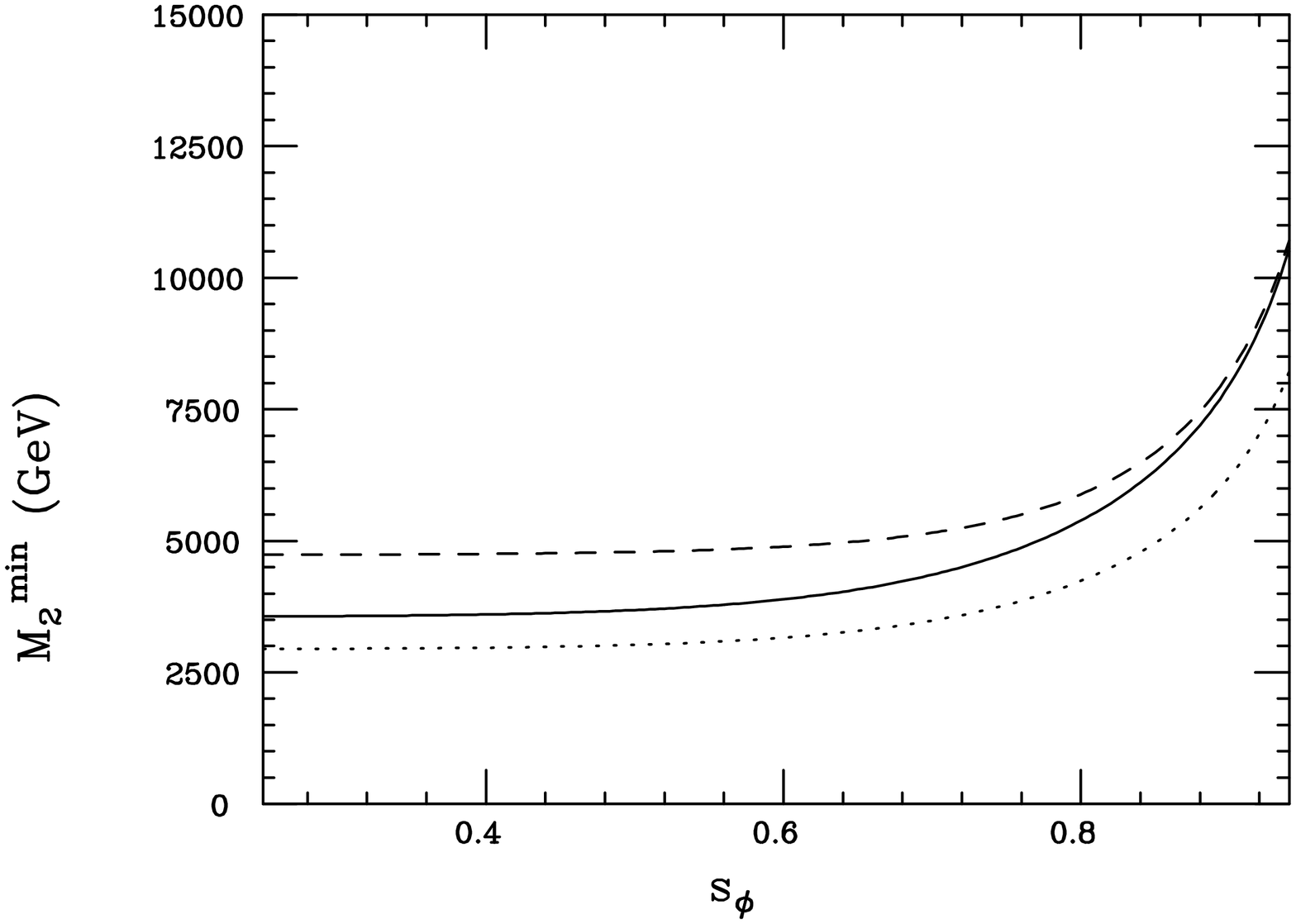,height=9.1cm,width=9.1cm,angle=-90}
\hspace*{-5mm}
\psfig{figure=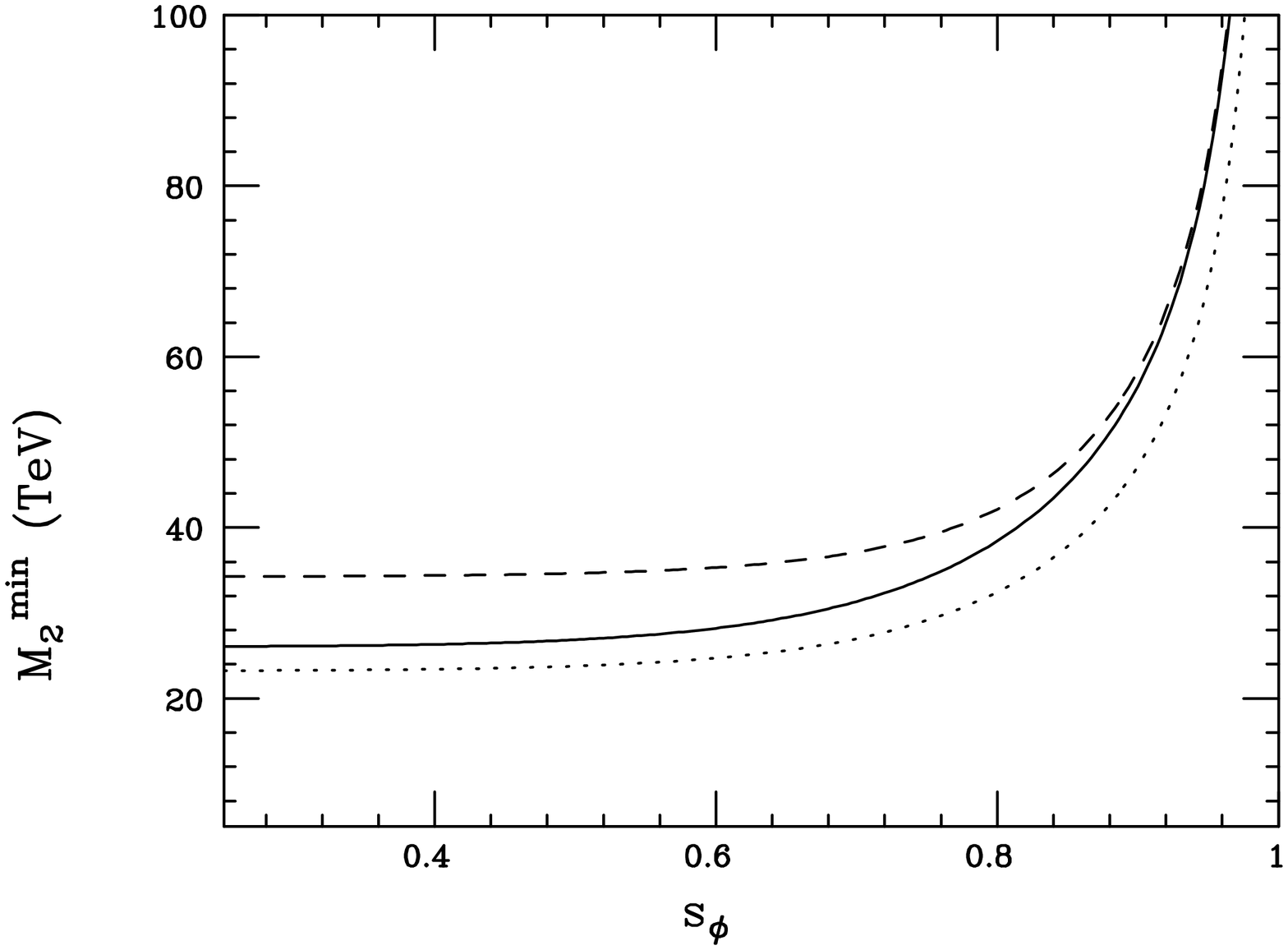,height=9.1cm,width=9.1cm,angle=-90}}
\vspace*{-0.6cm}
\caption{Indirect $Z'$ search reaches for the UUM at the 500 GeV NLC(left) 
and for a 5 TeV $e^+e^-$ NNLC collider(right) including initial 
state radiation. The dotted(solid, dashed) curve corresponds to the values 
obtained using leptonic(leptonic plus $b-$quark, all) observables. A luminosity 
of 50(1000) $fb^{-1}$ has been assumed for the NLC(NNLC).}
\label{fignlc}
\end{figure}
\vspace*{0.1mm}

In the analysis by Rizzo presented at this meeting{\cite {tgr}},  
the following standard set of observables were employed: 
$\sigma_{f}$, $A_{FB}^{f}$, $A_{LR}^{f}$, $A_{pol}^{FB}(f)$ 
where $f$ labels the fermion in the final state and, special to the case of 
the tau, $<P_\tau>$ and $P_\tau^{FB}$. Note 
that beam polarization plays an important role in this list of observables, 
essentially doubling its length. 
This was a first pass preliminary analysis wherein charged leptons as well as 
$b-$, $c-$, and $t-$quarks are considered simultaneously in obtaining the 
discovery reach. [Note: the results presented by Rizzo in the Snowmass workshop 
Subgroup summary talk did not include the $c$ and $t$ quark contributions.]
The basic approach follows that of Hewett and 
Rizzo{\cite {hr}} and is outlined in the review of Cvetic and 
Godfrey{\cite {rev}}, but now includes angular cuts, initial state 
radiation(ISR) in the $e^+e^-$ case but ignored for $\mu^+\mu^-$ collisions at 
the Large Muon Collider(LMC), finite identification efficiencies, systematics 
associated with luminosity and beam polarization($P$) uncertainties.  
For $e^+e^-$ colliders $P=90\%$ was assumed while for the LMC one can trade 
off a smaller effective $P$ through modifications{\cite {lmc}} in the 
integrated luminosity. The angular cuts, efficiencies, systematic errors, 
\etc, applied in all cases were assumed to 
be the same. This is probably extremely optimistic for the LMC since it is 
unclear whether a microvertex detector with suitable $b$ and $c$ 
identification efficiencies is possible in that collider environment.
Generically one find that ISR {\it lowers} the search reach 
by $15-20\%$ while beam polarization {\it increases} the reach by 
$15-80\%$ depending on the specific 
model and the machine energy, \ie, the increase is found to be smaller at 
larger values of $\sqrt s$.

\begin{table*}[htbp]
\leavevmode
\begin{center}
\label{$Z'$ nlc}
\begin{tabular}{lccccc}
\hline
\hline
Model  & NLC500  &NLC1000 &NLC1500 &NNLC 5 TeV& LMC 4 TeV \\
\hline
$\chi$  & 3.21 & 5.46 & 8.03 & 23.2 & 18.2 \\
$\psi$  & 1.85 & 3.24 & 4.78 & 14.1 & 11.1 \\
$\eta$  & 2.34 & 3.95 & 5.79 & 16.6 & 13.0 \\
I       & 3.17 & 5.45 & 8.01 & 22.3 & 17.5 \\
SSM     & 3.96 & 6.84 & 10.1 & 29.5 & 23.2 \\
ALRM    & 3.83 & 6.63 & 9.75 & 28.4 & 22.3\\
LRM     & 3.68 & 6.28 & 9.23 & 25.6 & 20.1 \\
UUM     & 4.79 & 8.21 & 12.1 & 34.7 & 27.3 \\
\hline
\hline
\end{tabular}
\caption{Indirect $Z'$ search reaches of lepton colliders in TeV employing 
all observables including the effects of ISR. The integrated luminosities of 
the NLC500, NLC1000, NLC1500, 
NNLC and LMC are assumed to be 50, 100, 100, 1000 and 1000 $fb^{-1}$, 
respectively.}
\end{center}
\end{table*}

Figure~\ref{fignlc} displays a set of sample results of this 
analysis at the 500 GeV NLC  and 5 TeV Next-to-Next Linear Collider(NNLC) for 
a $Z'$ of the UUM type. In particular, these plots show 
how the introduction of additional observables associated first with $b$ and 
then with $c$ and $t$ lead to an increased reach. Note that the inclusion of 
$c$ and $t$ in comparison to the leptons plus $b$ case leads to only a rather 
mild increase in the reach.  
Table~\ref{$Z'$ nlc} summarizes all these results for the search reaches of the 
various colliders for all of the above models. It is interesting to note that 
for the LMC the lack of significant ISR and the smaller 
polarization/luminosity are found 
to essentially cancel numerically in their affect on the $Z'$ search reach.

\begin{figure}[htbp]
\centerline{
\psfig{figure=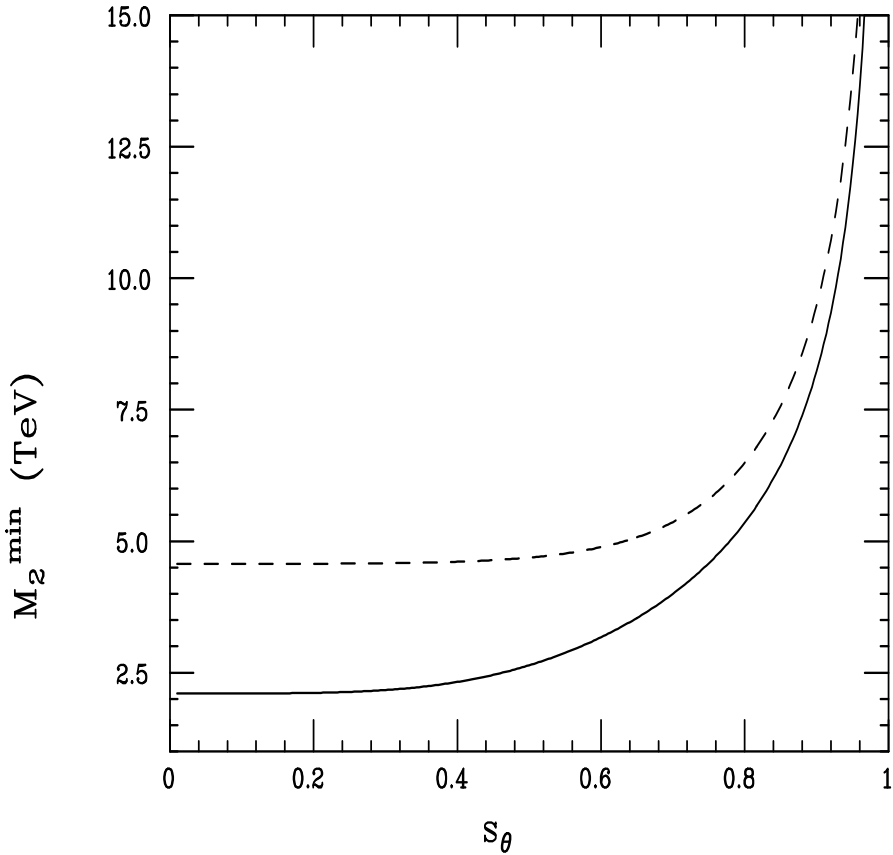,height=14cm,width=14cm,angle=-90}}
\vspace*{-0.9cm}
\caption{Search reach for the $Z'$ in topcolor models at the 500 GeV NLC 
with an integrated luminosity of 50 $fb^{-1}$. The solid line includes data 
from the $e,\mu,\tau$ and $b$ finals states; the dashed curve also includes 
data on $c$ and $t$.}
\label{ztc}
\end{figure}

It is possible to extend this technique to more exotic extended gauge 
models which do not obey family 
universality; a good example of this is the $Z'$ in topcolor-assisted 
technicolor models{\cite {tatc}} which is expected to lie below $\simeq 3$ 
TeV and above $\simeq 1.5$ TeV based on constraints from precision 
measurements{\cite {cat}}. The exact couplings depend upon a single free 
parameter, $s_\theta$. 
Fig.~\ref{ztc} shows that the search reach for this $Z'$ at the NLC is 
in excess of 4.7 TeV for all values of this parameter. Note the important 
role played by charm and top quark 
final states in obtaining this high reach. 

\vspace*{-0.5cm}
\nn
\vspace*{0.4mm}

\begin{figure}[htbp]
\leavevmode
\centerline{
\epsfig{file=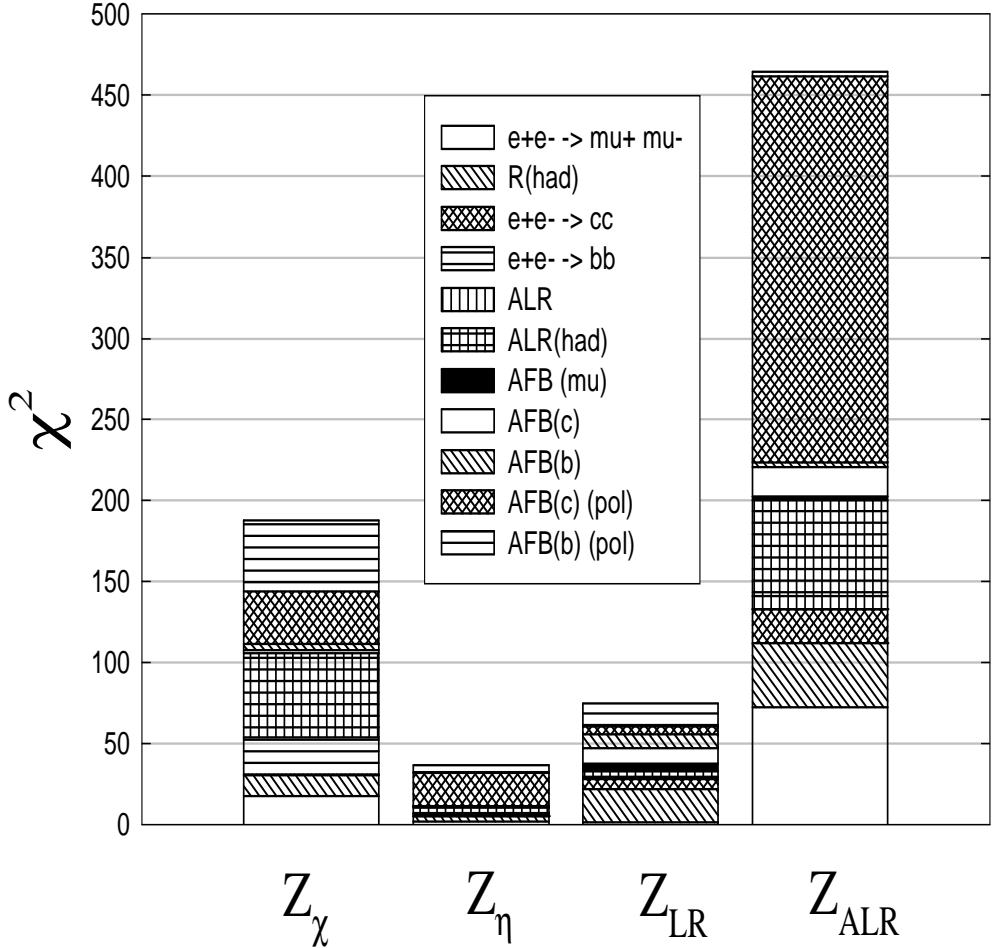,height=14cm,width=14cm,clip=}}
\vspace*{-0.3cm}
\caption{ Contributions to the total $\chi^2$ for a number of different 
observables used in the indirect $Z'$ searches in $e^+e^-$ colliders. The 
specific values are for the 500 GeV NLC with a luminosity of 50 $fb^{-1}$, 
$P=100\%$, and a $Z'$ mass of 2 TeV.}
\label{stevefig2}
\end{figure}

A parallel analysis of the capability of lepton colliders to indirectly 
discover a $Z'$ was performed by Godfrey{\cite {godfrey}} with a slightly 
different set of assumptions and observables, neglecting the effects of ISR. 
Numerically, the two analyses agree at the semi-quantitative level once the 
ISR contributions are taken into account. This is important in that it 
demonstrates that the $Z'$ search reach is not extremely sensitive to the 
detailed nature of the assumptions of a particular analysis as long as they 
are fairly reasonable. 
A very interesting part of Godfrey's analysis was a detailed examination 
of the various 
contributions which led to the $\chi^2$ used in setting the search reach. For 
the 500 GeV NLC, this is nicely displayed in Fig.~\ref{stevefig2} for four 
different $Z'$ models with the $Z'$ mass set to 2 TeV. The figure shows the 
variation in the size of the individual $\chi^2$ contributions is very 
significant. However, it also shows that the importance of the polarization 
asymmetries when information from various final state flavors are combined 
together.

\vspace*{-0.5cm}
\nn
\vspace*{0.4mm}

In principle the NLC can be run in the polarized $e^-e^-$ collision mode with a 
luminosity comparable to that for $e^+e^-$. Since both $e^-$ beams are 
polarized, the {\it effective} polarization is larger and, due to the large 
Moller cross section, there is significant sensitivity to the existence of a 
$Z'${\cite {fc}}. 
Unfortunately, an analysis of this situation including the 
effects of ISR was not available at the time of the meeting but a preliminary 
study by Cuypers{\cite {fc}} presented there indicated that the {\it ratio} 
of search reaches in the $e^+e^-$ and $e^-e^-$ modes might be stable under the 
modifications induced by ISR. 
Assuming this to be true, Rizzo{\cite {tgr}} thus repeated the previous 
$e^+e^-$ analysis {\it neglecting} ISR and also performed the 
complementary $e^-e^-$ analysis with the same cuts, efficiencies, \etc, ~and 
then took the ratio of the resulting reaches for a given extended gauge model. 
The results of this analysis for NLC500 are shown in 
Table~\ref{$e^-e^-$}. Here we see that in general the $e^-e^-$ reach is 
superior to that obtained in the $e^+e^-$ mode when only the leptonic 
final states are used, 
consistent with the results obtained in Ref.{\cite {fc}}. However, as soon as 
one adds the additional information from the quark sector, $e^+e^-$ regains 
the lead in terms of $Z'$ mass reach. Combining the leptonic and quark data 
together in the $e^+e^-$ case always results in a small value for the ratio.

\begin{table}[htbp]
\begin{center}
\label{$e^-e^-$}
\begin{tabular}{lccc}
\hline
\hline
Model  & $\ell$  & $\ell+b$ & $\ell+b,c,t$ \\
\hline
$\chi$  &  1.10 & 0.900 & 0.896 \\
$\psi$  &  1.20 & 0.711 & 0.673 \\
$\eta$  &  1.07 & 0.813 & 0.650 \\
I       &  1.06 & 0.813 & 0.813 \\
SSM     &  1.30 & 0.752 & 0.667 \\
ALRM    &  1.20 & 1.12  & 0.909 \\
LRM     &  1.02 & 0.483 & 0.432 \\
UUM     & 0.891 & 0.645 & 0.496 \\
\hline
\hline
\end{tabular}
\caption{Ratio of $e^-e^-$ to $e^+e^-$ indirect $Z'$ search reaches at a 500 
GeV NLC with an integrated luminosity of 50 $fb^{-1}$ in either collision 
mode. ISR 
has been ignored. The columns label the set of the final state fermions used in 
the $e^+e^-$ analysis.}
\end{center}
\end{table}

\begin{figure}[htbp]
\leavevmode
\centerline{
\input{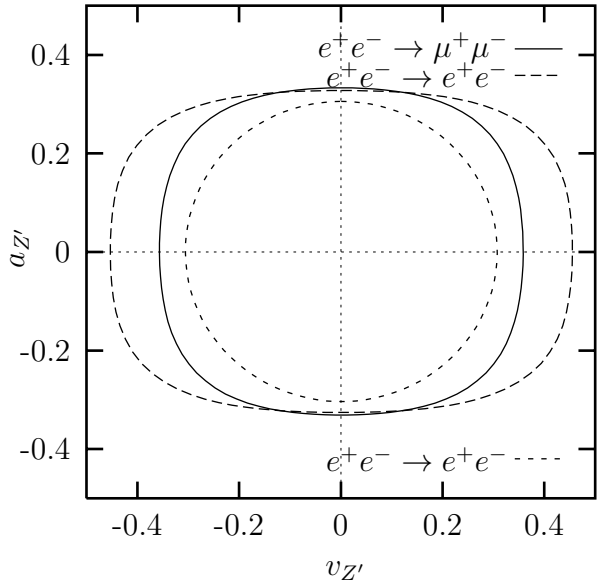}}
\vspace*{-0.3cm}
\caption{Contours of observability at 95$\%$ CL for the reduced $Z'$ couplings 
including the effects of ISR, polarization and luminosity uncertainties, as 
well as the angular resolution of the detector. These results are for a 500 
GeV NLC with $P=90\%$ with a luminosity of 50(25) $fb^{-1}$ in the 
$e^+e^-(e^-e^-)$ mode.}
\label{cuypersfig}
\end{figure}

Of course, we need to verify these results directly; in a contribution to 
these proceedings, Cuypers examined the influence of a number of systematic 
effects in the searches for $Z'$'s in the purely leptonic processes 
$e^+e^- \to \mu^+\mu^-$ as well as in Bhabha and Moller 
scattering{\cite {cuy}}. He has now demonstrated that for these processes the 
effects of ISR modify the $Z'$ search reaches by essentially the same 
amount $\simeq 15\%$. 
Cuypers also showed that the systematic uncertainties in both beam 
polarization (since both beams are polarized) and angular resolution (due to 
the $t$-channel pole) are far more important in Moller scattering than in 
$e^+e^- \to \mu^+\mu^-$. In fact, for Bhabha scattering, Cuypers has found 
that the angular resolution is the largest source of systematic error. 
Including all systematic effects, Bhabha scattering was found to be the least 
sensitive to the existence of a $Z'$. A comparison of the sensitivities 
of these three processes to a new $Z'$ at a 
500 GeV NLC with $P=90\%$ is shown in Fig.~\ref{cuypersfig}.

\vspace*{-0.5cm}
\nn
\vspace*{0.4mm}

  A $W'$ can also be produced in pairs in $e^+e^-$ annihilation via 
$s$-channel $\gamma,Z,Z'$ exchanges and some model-dependent $t$-channel 
exchange. For example, in the LRM(UUM), a heavy right-handed (massless 
left-handed) neutrino  is exchanged in the $t$-channel. While the cross 
sections for this process are large{\cite {wrref1}}, the kinematic reach for 
direct production is rather poor $\leq \sqrt s/2$. In the LRM,  
it is also possible to produce like-sign $W_R$ pairs in $e^-e^-$ collisions if 
the right-handed neutrino is a Majorana particle{\cite {wrref2}}. Of course, 
the reach is the same as in $e^+e^-$ collisions. One possible way to extend 
the direct discovery range  
is to produce one on-shell and one off-shell $W'${\cite {wrref3}}. 
In this case $W'$ masses as large as $\simeq 0.8\sqrt s$ can be reached. 
Another possibility is to employ the $\gamma e$ collision mode where the $W'$ 
is produced in association with some other fermion; in the LRM case this 
pushes the reach almost up to the kinematic limit: $m_{W_R}+m_N \leq \sqrt s$, 
where $m_N$ is the mass of the right-handed neutrino{\cite {wrref4}}.
It is clear from this discussion that for a more massive $W'$, we need to 
perform an indirect search as has just been discussed in the case of a $Z'$.

Since virtual $W'$'s are not conventionally exchanged in $e^+e^-\to f\bar f$ 
processes, it is difficult to obtain indirect mass limits. One possibility, 
explored by Hewett{\cite {joa}} in a first pass analysis for these 
proceedings, is the famous 
`neutrino-counting' process $e^+e^-\to \nu \bar \nu \gamma$. In the SM, this 
reaction proceeds though the `subprocess' $e^+e^-\to \nu \bar \nu$, which 
occurs via $s-$channel $Z$ and $t-$channel $W$ exchanges and an additional 
photon is then allowed to be emitted by any charged leg. In models with new 
$W'$ and $Z'$ gauge bosons there will be additional graphs that can lead to 
modifications in the SM result. For a given $W'$ mass, the corresponding 
$Z'$ mass is fixed by a model dependent relationship as discussed above. The 
SM $W$ and $W'$ are treated as contact interactions in this first 
approximation. Thus in the LRM (assuming Dirac neutrinos) or the 
UUM we need only specify 
$\kappa$ or $s_\phi$ as well as $M_{W'}$ to perform the complete calculation 
if we neglect any possible mixing among the gauge bosons. 
Unfortunately, this radiative process is suppressed in 
comparison to the usual fermion pair rate by an additional power of $\alpha$ 
as well as by three-body phase space, though these are somewhat offset by 
the appearance of large logarithms. We might thus 
expect that the available statistical power may not be able to provide much 
of a search reach, but it is clear that any extension 
beyond $M_{W'}\geq \sqrt s$ is important. 

To render the process observable 
(and also to make the cross section finite by removing infrared and colinear 
divergences), the photon 
energy is assumed to be $\geq 0.05\sqrt s$ and to make an angle with the 
electron or positron beam 
directions $\geq 20^\circ$, which should be well inside the NLC detector. 
What observables are useful in obtaining constraints? In addition to the total 
cross section, we can form the Left-Right asymmetry, $A_{LR}$, using the 
initial beam polarization. In the SM, the value of $A_{LR}$ is close to unity 
due to the rather strong influence of the $W$. Unfortunately, for interesting 
$W'$ masses this situation is not altered and one finds that $A_{LR}$ is not 
useful. One can also, in principle, use the energy and angular 
distributions of the final state photon; however, a short analysis 
demonstrates the the by far dominant influence here is just QED in the $W/W'$ 
contact interaction approximation. We are thus 
left with only the total cross section as the only useful observable.

\vspace*{-0.5cm}
\nn
\begin{figure}[htbp]
\centerline{
\psfig{figure=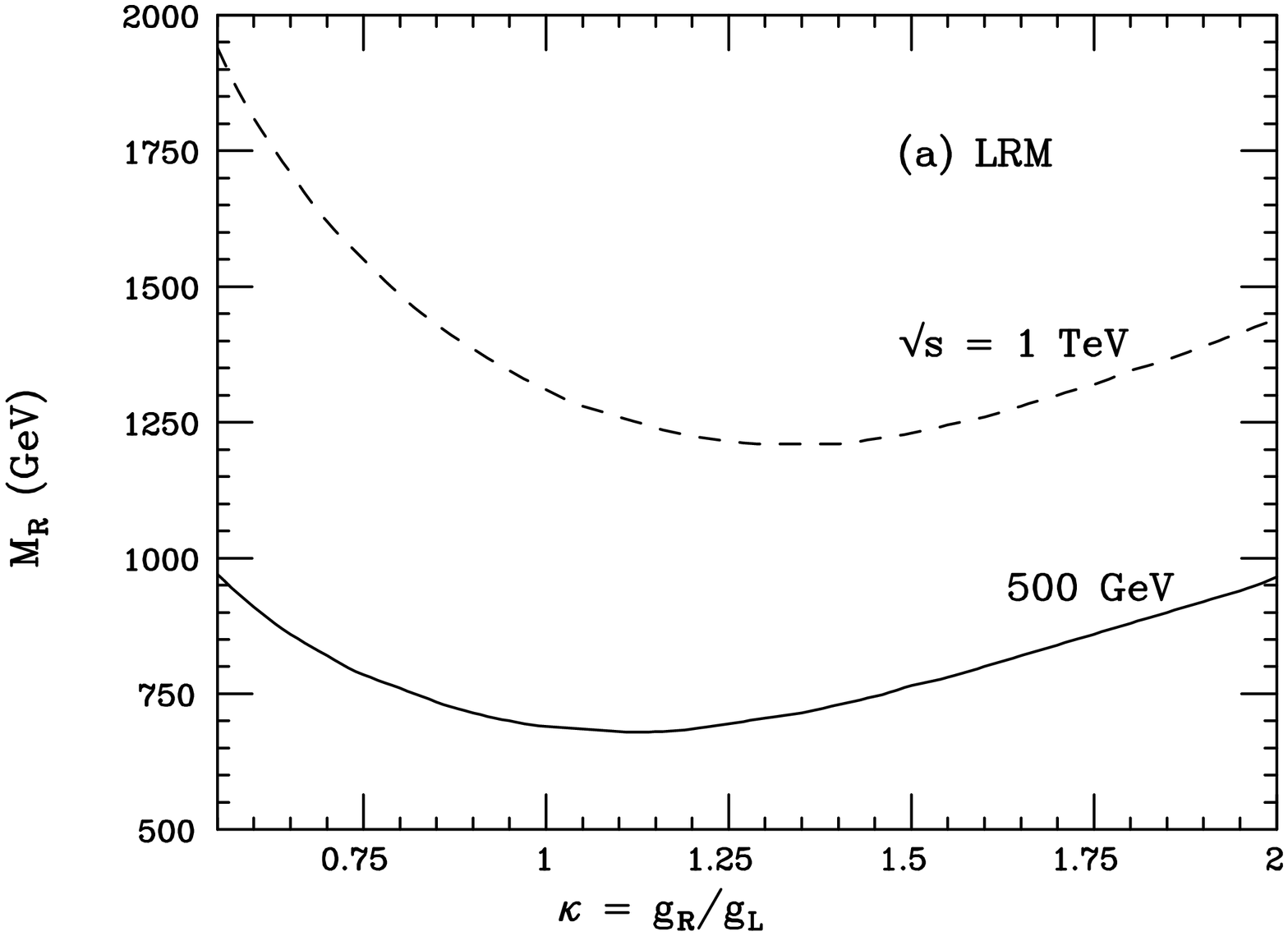,height=9.1cm,width=9.1cm,angle=-90}
\hspace*{-5mm}
\psfig{figure=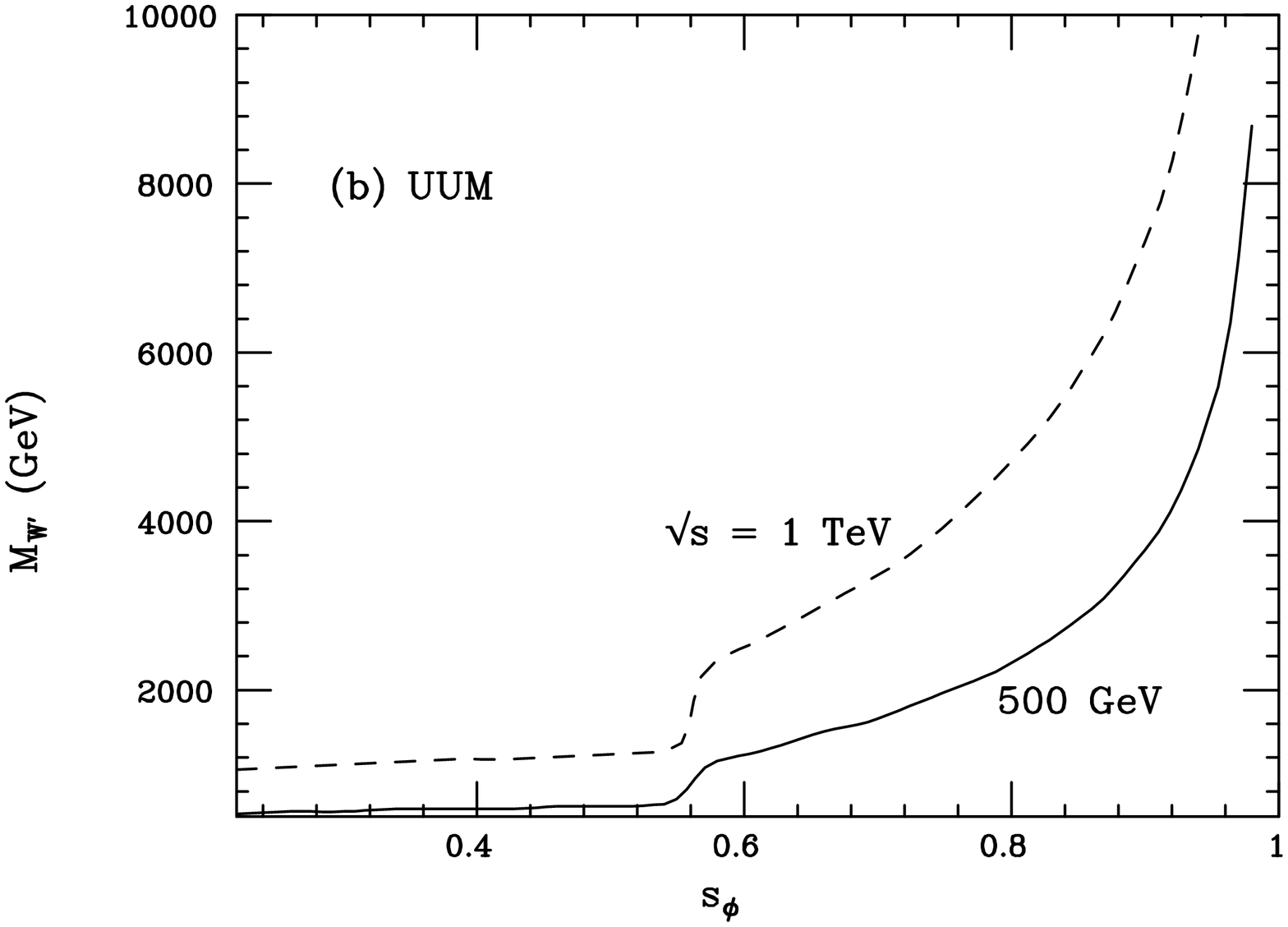,height=9.1cm,width=9.1cm,angle=-90}}
\vspace*{-0.6cm}
\caption{$95\%$ CL lower bound on the $W'$ in (a) the LRM as a function of 
$\kappa$ and (b) the UUM as a function of $s_\phi$. In each case the 
lower(upper) curve corresponds to a center of mass energy of 500 GeV(1 TeV) 
and an integrated luminosity of 50(200) $fb^{-1}$.}
\label{hewett}
\end{figure}
\vspace*{0.1mm}

The results of Hewett's analysis for the exclusion reach of this process for 
a new $W'$ can be seen in Fig.~\ref{hewett} for both the LRM and the UUM 
cases. This figure show the minimum value of the $W'$ mass as a function of 
either $\kappa$ or $s_\phi$ for NLC collider energies of 500 GeV and 1 TeV 
and luminosities of 50 and 200 $fb^{-1}$ respectively. For the LRM case, the 
limits range from $\simeq 680$ GeV to $\simeq$ 975 GeV above the kinematic 
limit in the $\sqrt s$=500 GeV case for $0.55\leq \kappa \leq 2$. 
For the case of a 1 TeV collider the corresponding reach ranges from 1200 to 
1950 GeV. For the UUM with small values of $s_\phi$, the reach is found to not 
be significantly greater than $\sqrt s$. As $s_\phi$ grows beyond 0.5, the 
leptonic couplings of the $W'$ and $Z'$ increase and the reach increases 
dramatically to several times $\sqrt s$ for both the 500 GeV and 1 TeV NLC.
For both models we see that reasonable exclusion reaches are obtainable.
The influence of the contact interaction approximation will be examined in a 
future analysis{\cite {joa2}}.

\section{Extraction of Coupling Information}

Once a new gauge boson is found a new era begins, \ie, to ascertain all of its 
properties. Only if we know as much as possible about the new $Z'/W'$ will we 
be able to determine its origin within a more general extended gauge model. 
Both hadron and lepton colliders can play important and 
complementary roles in reaching this goal. Each has its own strength and 
weaknesses and are discussed separately.

\subsection{Hadron Colliders}

The determination of the couplings of a $Z'$ at a hadron collider is a highly 
non-trivial task due to both large backgrounds and limited statistics. In our 
discussion below, we focus on the determination of $Z'$ couplings to the SM 
fermions at the LHC. Certainly the same problems are to be faced at other 
hadron colliders. 
The recent review of $Z'$ physics by Cvetic and Godfrey{\cite {rev}} 
shows that in an 
idealized world, without backgrounds or systematic errors to worry about, 
the LHC will be able to do a reasonable job at extracting the couplings of a 
new $Z'$ if its mass is not too much greater than 1 TeV by combining a series 
of different measurements in a simultaneous fit. What we really want 
to know is how well this program can be performed by a  real LHC detector.

At first glance it would appear that statistics should {\it not} be a problem 
at, \eg, the LHC with a luminosity of $100 fb^{-1}$, but this is not 
always true. 
While the typical search reach for a 
$Z'$ at the LHC is near 5 TeV this would give us only a few events. To even 
begin to {\it analyze} a $Z'$ requires more than 100 events in the discovery 
channel. This tells us that it is unlikely that we will ever gain sufficient 
information about a $Z'$ much heavier than about 3-3.5 TeV{\cite {wulz}} 
unless it had a particularly large production cross section or significantly 
more luminosity were to be available. In reality, the reach for coupling 
analysis is far inferior to the 3-3.5 TeV range at the LHC.

\begin{figure}[htbp]
\centerline{
\psfig{figure=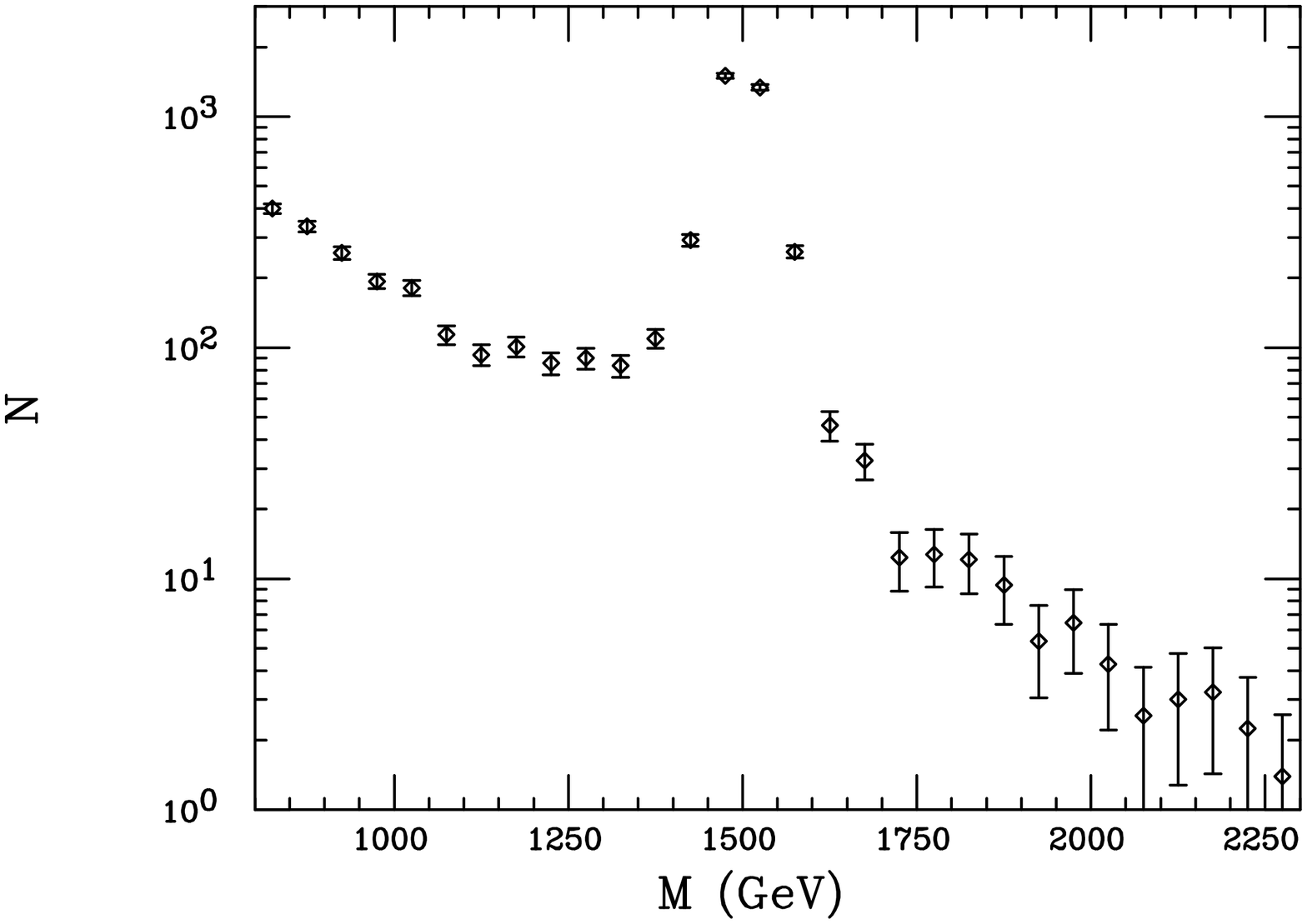,height=9.1cm,width=9.1cm,angle=-90}
\hspace*{-5mm}
\psfig{figure=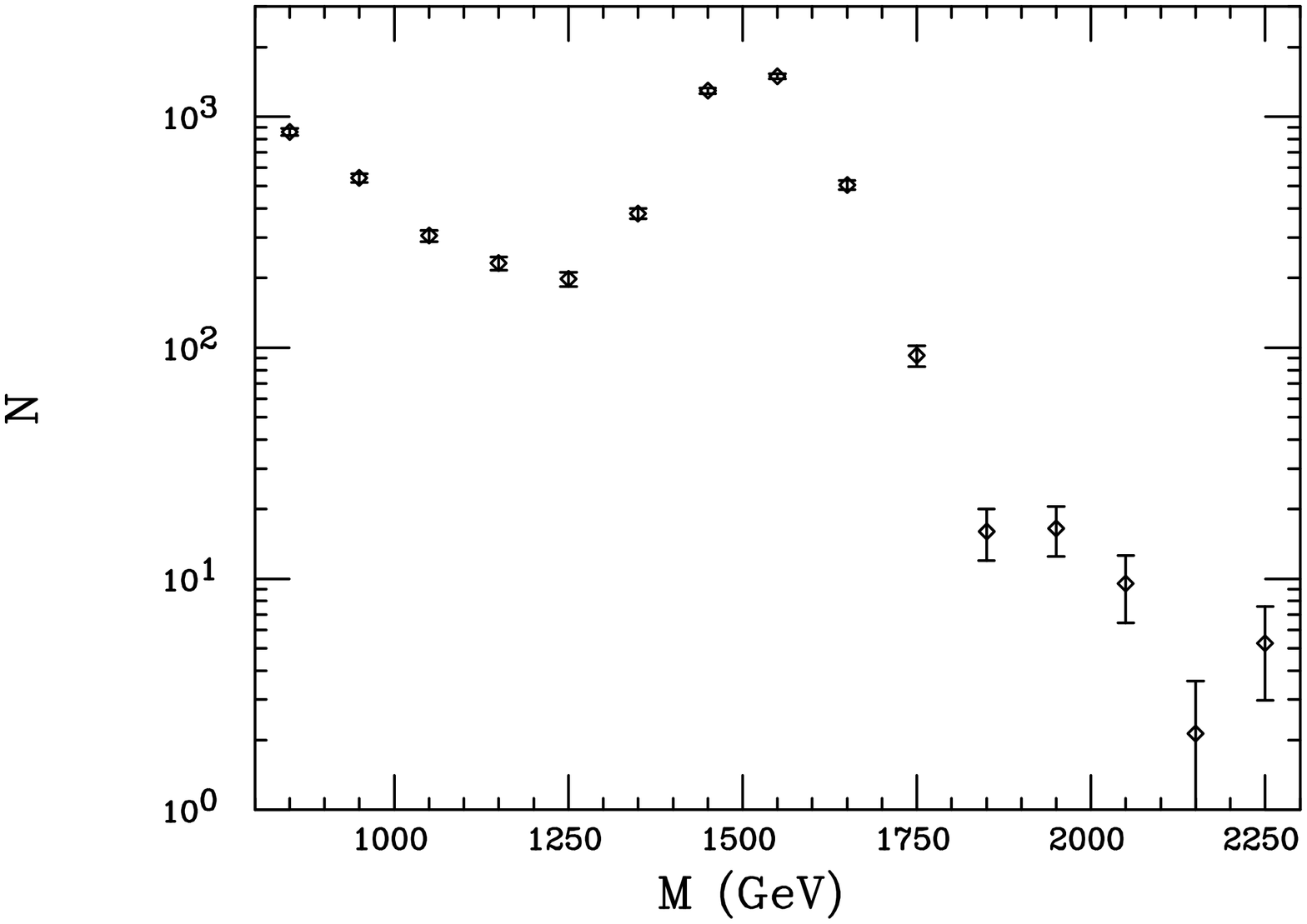,height=9.1cm,width=9.1cm,angle=-90}}
\vspace*{-0.6cm}
\caption{Simulation of a typical $Z'$ lepton pair invariant mass distribution 
assuming $M_{Z'}$=1.53 TeV for electrons(left) and muons(right) 
smeared with the ATLAS(CMS) resolutions at the LHC assuming a luminosity of 
$100fb^{-1}$ and $|\eta_l|\leq 2.5$. The bin size is 50(100) GeV; only 
Drell-Yan backgrounds are included.}
\label{sigbin}
\end{figure}

When a $Z'$ is discovered, both ATLAS and CMS will easily measure its mass, 
total width($\Gamma_{tot}$), and its production cross section in the leptonic 
channel($\sigma_l$), which in the narrow width approximation is given by 
$\sigma(q\bar q \to Z')B_l(=\Gamma(Z'\to \ell^+\ell^-)/\Gamma_{tot})$. 
Unfortunately, 
this last observable {\it cannot} be used to extract coupling information 
since the value of $B_l$ depends not only on the conventional quark and lepton 
couplings to the $Z'$(which we want to determine) but also on possible decays 
to SUSY partners, exotic states, \etc. Fortunately, however, the product 
$\sigma_l\Gamma_{tot}$ is decay mode independent and will tell us something 
about the overall $Z'$ coupling strength. Of course, the production of a $Z'$ 
at the LHC in the real world does not 
look like the narrow width approximation but more like Fig.~\ref{sigbin}, so 
that resolution effects need to be deconvoluted and efficiencies and 
backgrounds accounted for 
before this product of observables can be readily determined. 

This observation reminds us 
that past analyses of the extraction of $Z'$ coupling information at hadron 
colliders have not accounted for detector issues and have systematically 
relied on the narrow width approximation. (It is also generally assumed 
that there will be 
little uncertainty due to variations in the parton densities. This may be a 
valid assumption in 10 years time when the LHC begins analyzing data!) 
Before a $Z'$ is found at the LHC 
we need to revisit these older analyses and try to understand how well the 
proposed observables can be measured in a more realistic 
situation. We began this exercise at the Snowmass workshop and report below on 
some of our results and observations. A complete model-independent 
coupling extraction analysis through the use of detector simulations for 
the LHC is still some years away from being demonstrated.

\vspace*{-0.5cm}
\nn
\vspace*{0.1mm}

Since the lepton-pair channel is the discovery channel for a $Z'$, it is 
obvious that we should try to extract as much information as possible there. 
Several observables have been proposed{\cite {old}}:
\begin{itemize}
\item The forward-backward asymmetry, $A_{FB}$;
\item The rapidity ratio, $r_{y1}$, the ratio of cross sections in the central 
rapidity region in comparison to larger rapidities;
\item The $\tau$ polarization asymmetry, $A_\tau$, in $Z'\to \tau^+\tau^-$;
\item The various polarization asymmetries that can be formed if at least one 
polarized proton beam is available. Clearly, this possibility also relies on 
having excellent knowledge of the polarized parton densities of the proton at 
$Q^2 \simeq 1$ TeV$^2$. It would seem that such observables will not be used in 
the first round effort to disentangle $Z'$ couplings. 
\end{itemize}
Note that all these observables are {\it ratios} of cross sections and are thus 
less subject to systematic uncertainties and are also independent of the $Z'$ 
decay modes. Since we are assuming that hundreds of $Z'$ events are available 
the measurements of these observables are not statistics limited. In 
principle, we would like to have available Monte Carlo studies of each of 
these quantities including detector simulations. This work was initiated 
during the workshop. 

\begin{figure}[htbp]
\centerline{
\psfig{figure=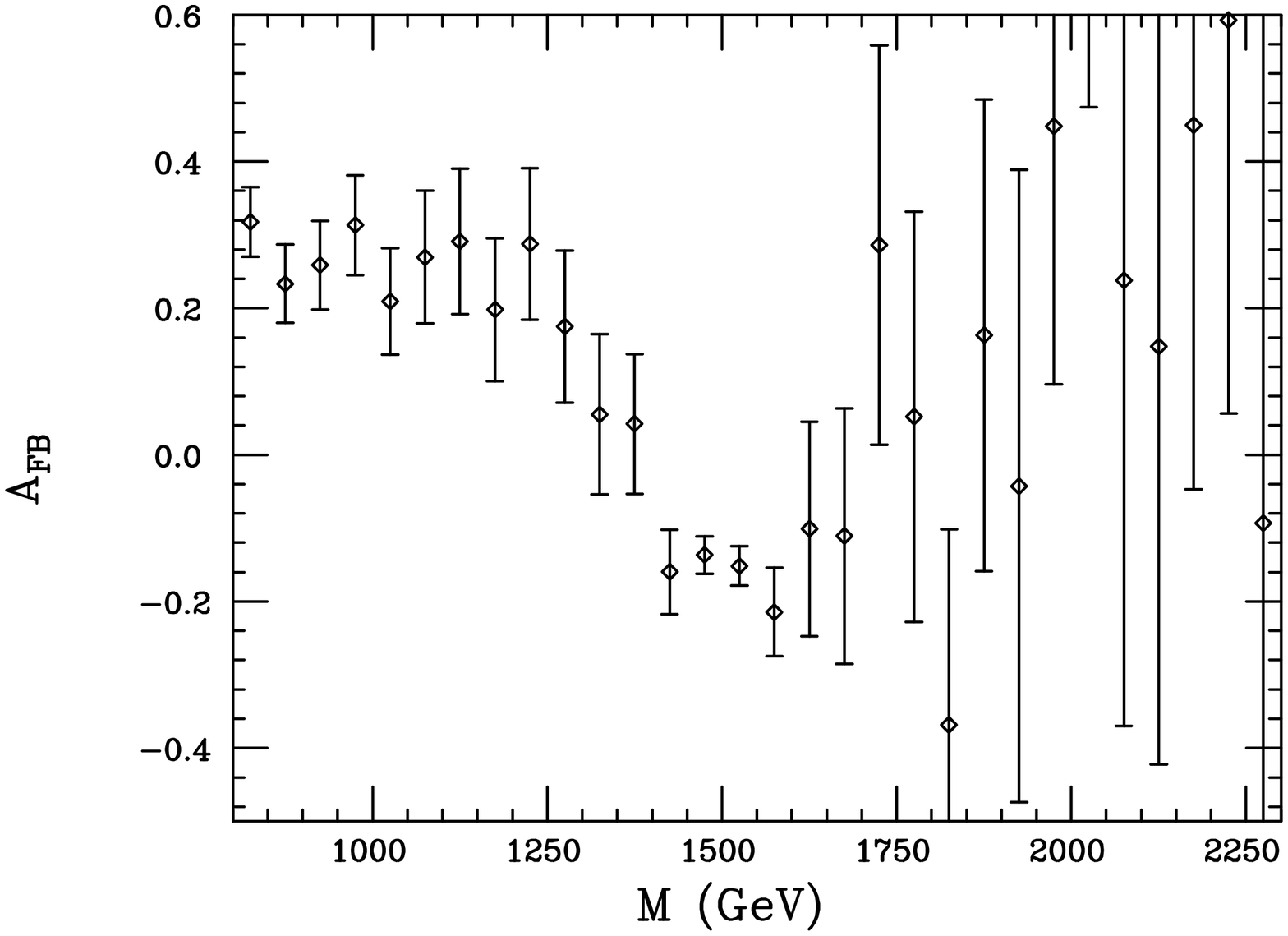,height=9.1cm,width=9.1cm,angle=-90}
\hspace*{-5mm}
\psfig{figure=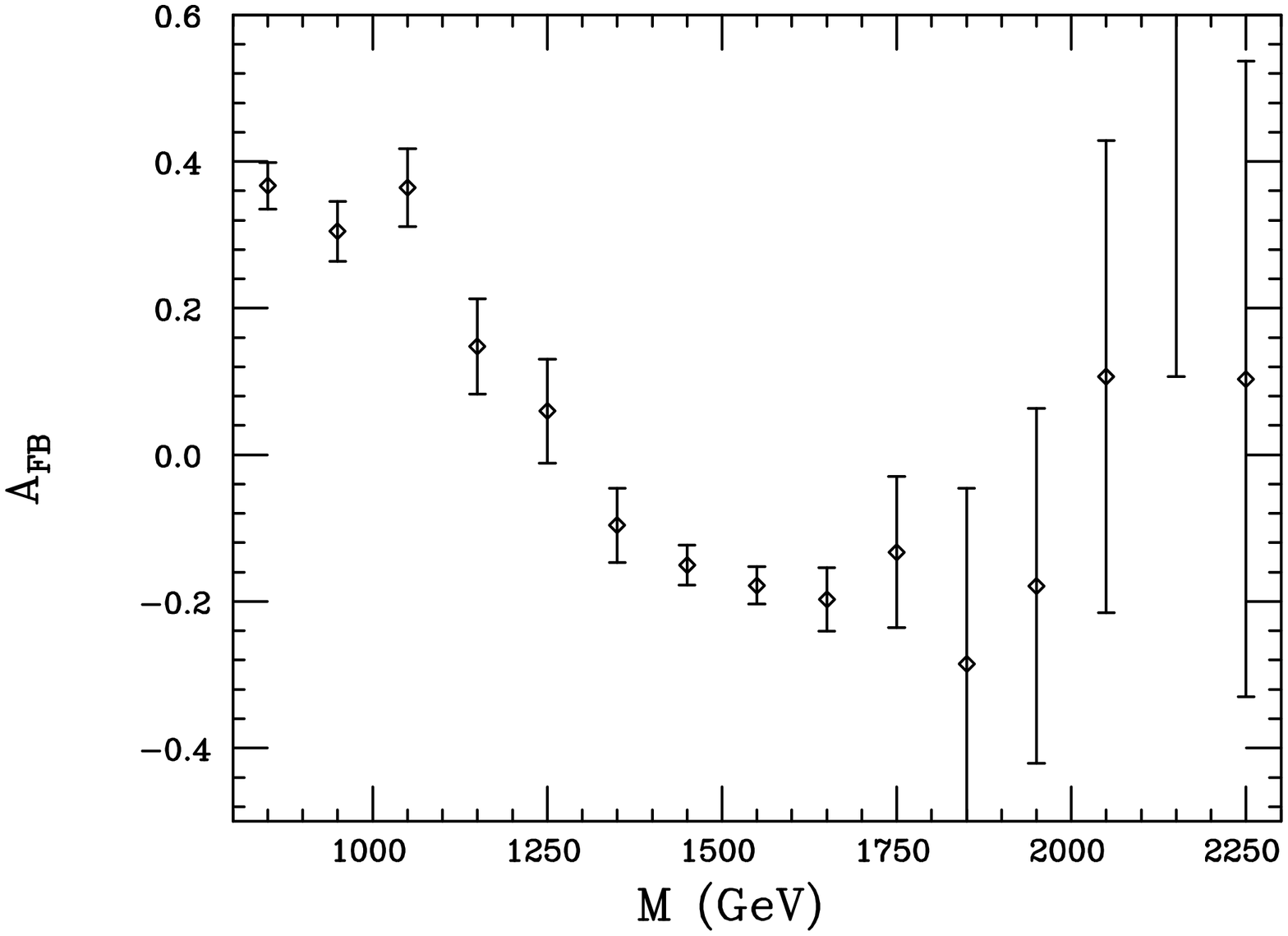,height=9.1cm,width=9.1cm,angle=-90}}
\vspace*{-0.6cm}
\caption{Simulation of the $A_{FB}$ for a typical $Z'$ with a mass of 
$M_{Z'}$=1.53 TeV for electrons(left) and muons(right) 
smeared with the ATLAS(CMS) resolutions at the LHC assuming a luminosity of 
$100fb^{-1}$ and $0.3\leq |\eta_l|\leq 2.5$. The bin size is 50(100) GeV; only 
Drell-Yan backgrounds are included. The low $\eta$ region is removed to 
eliminate ambiguous hemisphere assignments for the leptons.}
\label{asymbin}
\end{figure}

$A_{FB}$ is perhaps the most well-studied of this set of observables for 
purposes of coupling extraction but again generally only in the narrow 
width limit. 
Unfortunately, as a function of the dilepton mass, $A_{FB}$ will look more 
like Fig.~\ref{asymbin} when it is first measured and not a simple number as 
given by the narrow width estimate. 

\vspace*{-0.5cm}
\nn
\vspace*{0.1mm}

\begin{figure}[htbp]
\leavevmode
\centerline{
\epsfig{file=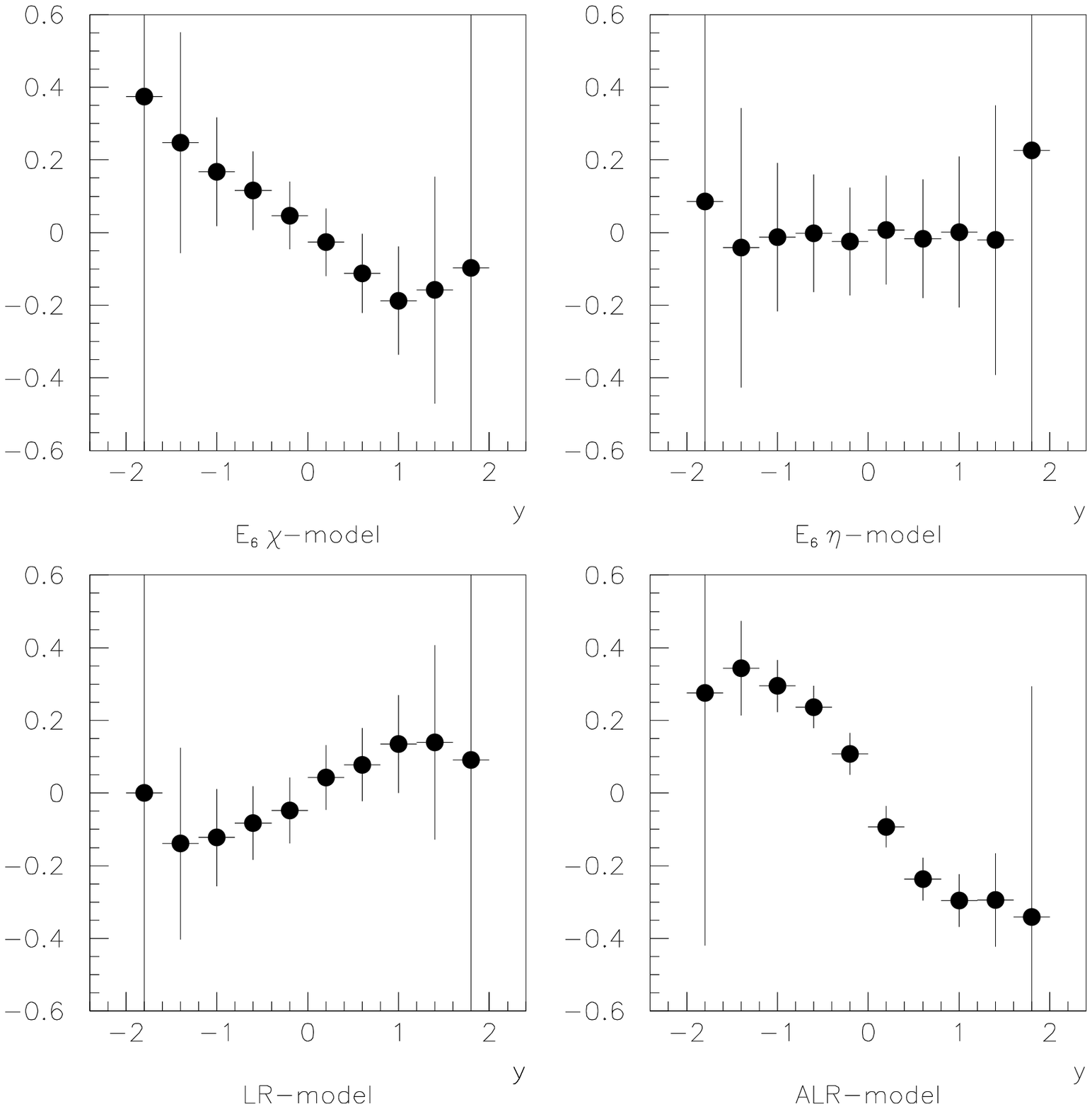,height=14cm,width=14cm,clip=}}
\vspace*{-0.3cm}
\caption{Simulation of the $Z'$ forward-backward asymmetries for 
different models as a function 
of rapidity($y$) assuming $M_{Z'}=2$ TeV as seen by the CMS detector in the 
dimuon channel. Signal and background have been integrated over the lepton 
pair mass range $M_{Z'}\pm \Gamma_{tot}$.}
\label{claudia1}
\end{figure}

\begin{figure}[htbp]
\leavevmode
\centerline{
\epsfig{file=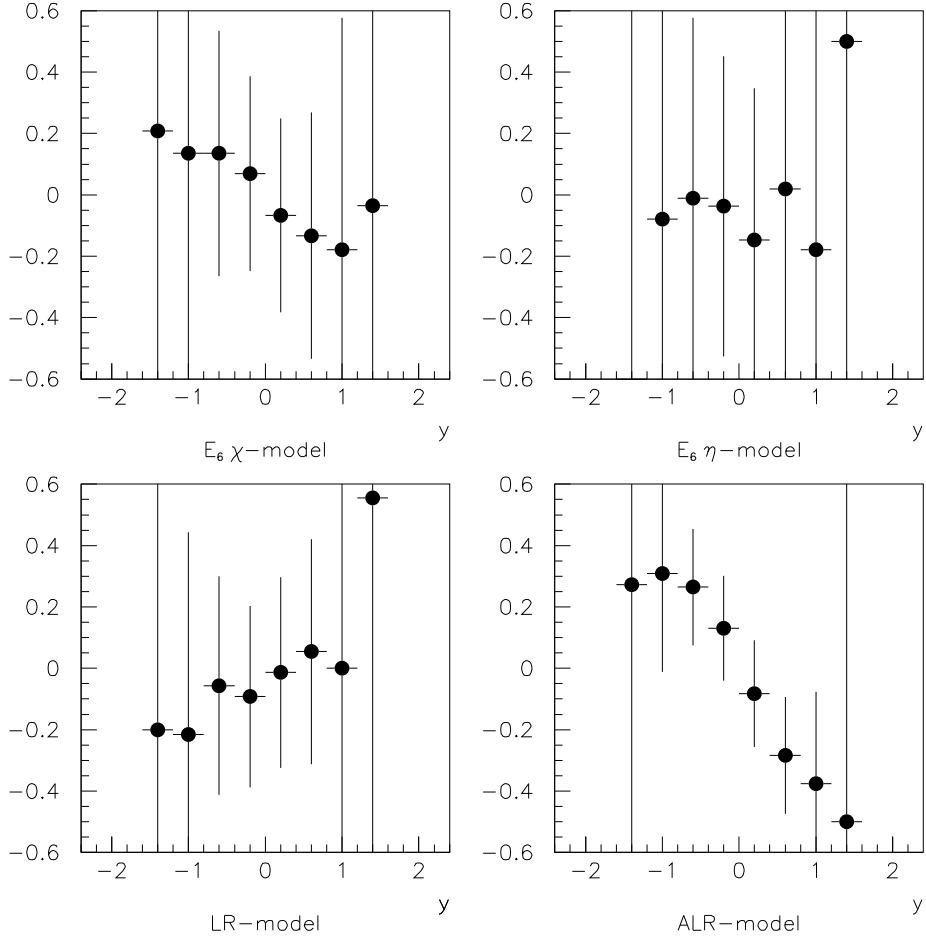,height=14cm,width=14cm,clip=}}
\vspace*{-0.3cm}
\caption{Same as the previous figure but now for $M_{Z'}=3$ TeV.}
\label{claudia2}
\end{figure}

\begin{figure}[htbp]
\centerline{
\psfig{figure=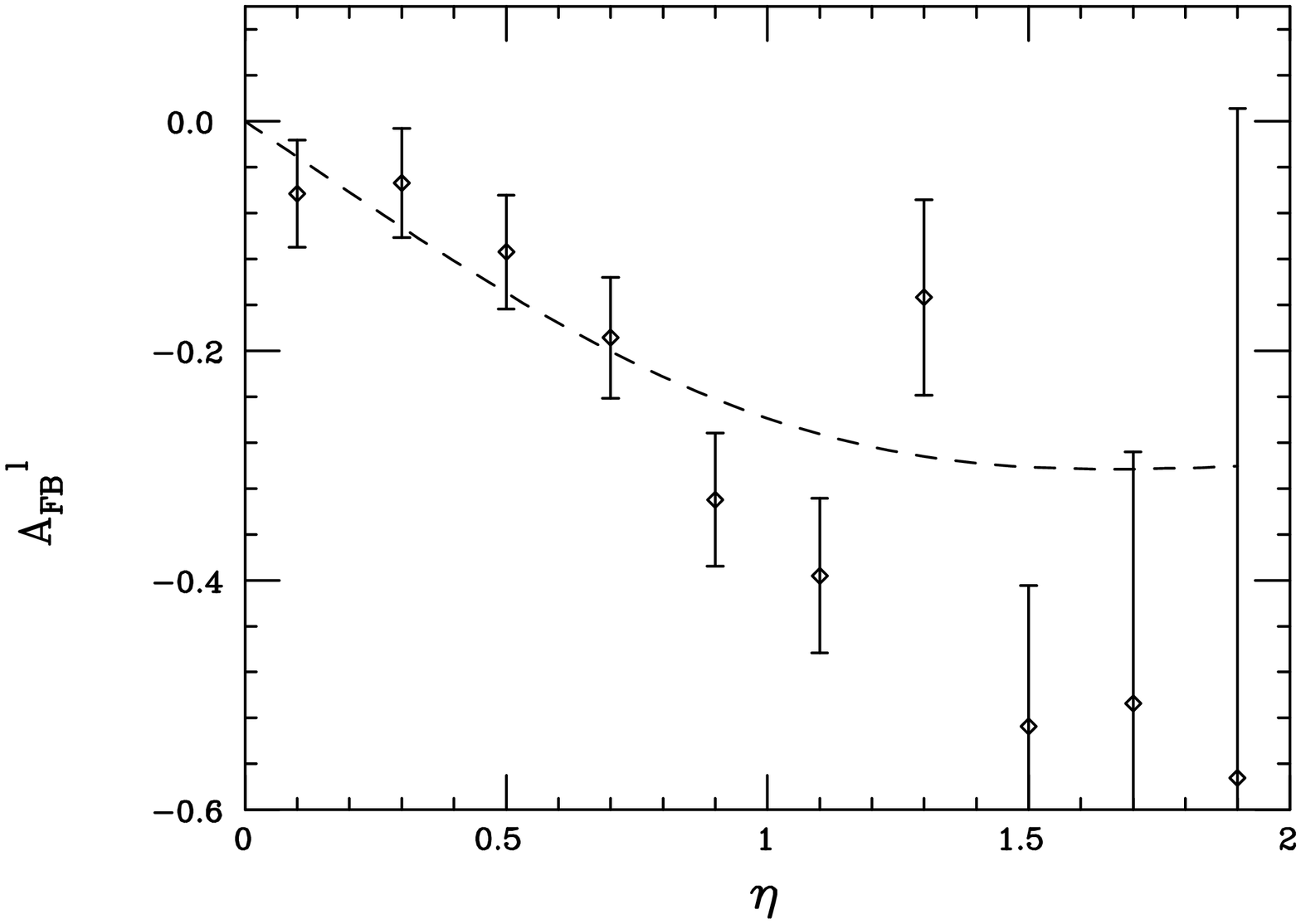,height=9.1cm,width=9.1cm,angle=-90}
\hspace*{-5mm}
\psfig{figure=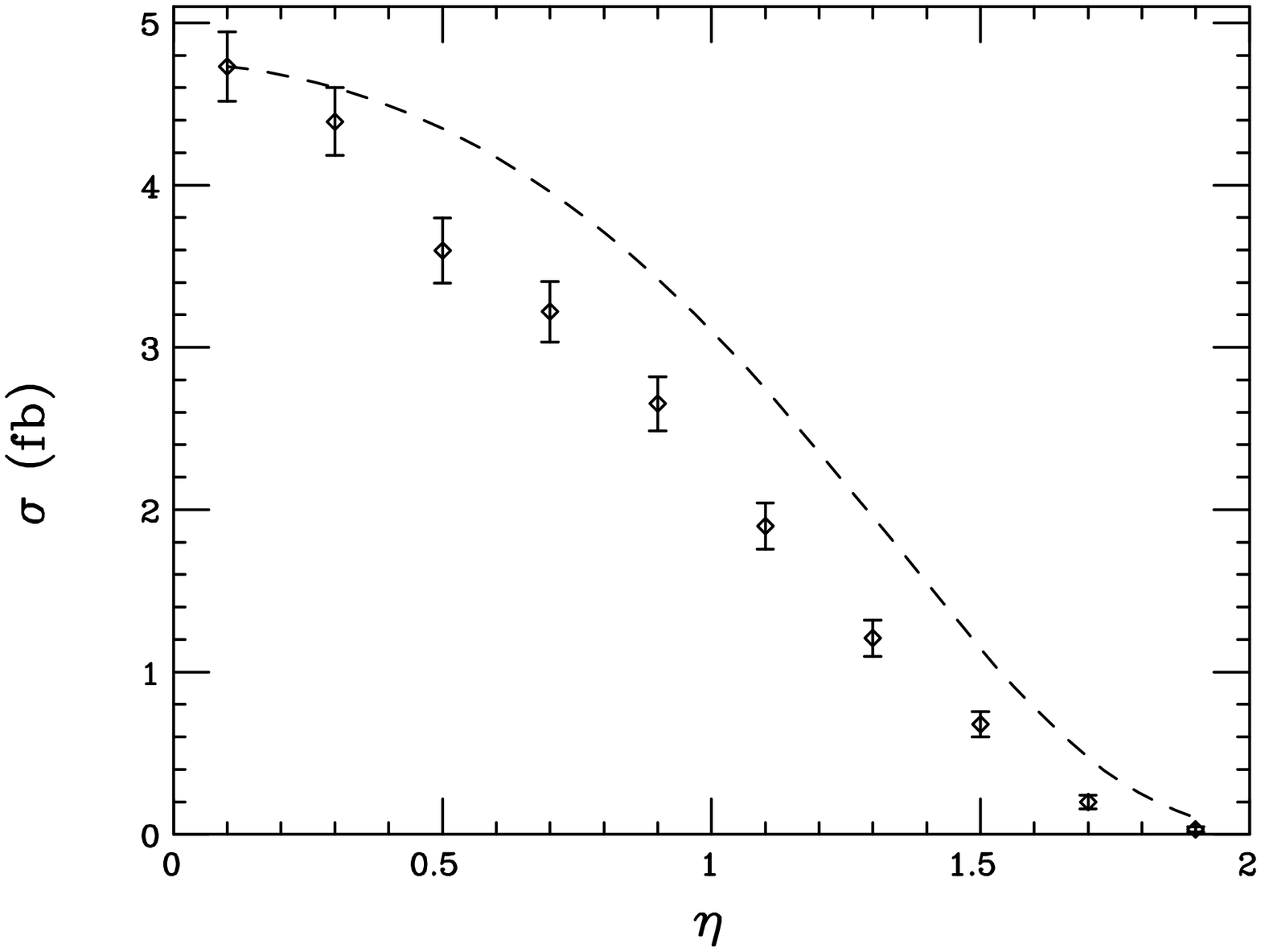,height=9.1cm,width=9.1cm,angle=-90}}
\vspace*{-0.6cm}
\caption{A typical comparison of narrow width approximation expectations with 
a simulation for an ATLAS-like detector in the $Z'\to e^+e^-$ mode 
assuming a luminosity of $100 fb^{-1}$ at the LHC. The $Z'$ mass is 1.53 TeV. 
On the left(right) is $A_{FB}(\sigma)$ as a function of rapidity. The dashed 
curve is the narrow width result which has been rescaled in the $\sigma$ case 
to go through the first data point. Only events in the mass bin 
$M_{Z'}\pm \Gamma_{tot}$ are included.}
\label{newdis}
\end{figure}

To proceed one needs to cut away as much of the underlying Drell-Yan 
background as possible without too much of a loss in statistics. A mass 
cut such as $M_{Z'}\pm (1-2)\Gamma_{tot}$ is found to be most useful. 
For the sample model in Fig.~\ref{sigbin}, a cut of $\pm (2)\Gamma_{tot}$ 
captures about 60(72)$\%$ of the $Z'$ with a background contamination of 
less than about $2\%$ for electron pairs. (For wider $Z'$'s, as well as for 
muon pairs, the backgrounds could be 
significantly worse and tighter invariant mass cuts should be applied.) 
The events remaining after this cut 
can then be plotted vs. rapidity as has been done in the analysis of 
Wulz{\cite {wulz}} with the full CMS detector simulation. 
It is clear from Figures~\ref{claudia1} and ~\ref{claudia2}, that the 
$Z'$'s are reasonably distinguishable even with a mass of 3 TeV. However, it is 
not so easy to go from {\it real data} that may look like these plots to the 
extraction of coupling information. (Remember that we want to do more than 
distinguish models, we want to get at the $Z'$ couplings.) As before, 
resolutions can be deconvoluted, but the background's contribution to the 
asymmetry may be potentially large. Numerically, however, the narrow width 
approximation works fairly well in practise and gives reasonable results at 
the level of $10-15\%$ for both the rapidity-integrated asymmetry as well as 
the dependence of $A_{FB}$ on rapidity. Fig.~\ref{newdis} shows a direct 
comparison between Monte Carlo $A_{FB}$ `data' generated using a simplified 
simulation of the ATLAS 
detector and the narrow width approximation expectations for a typical $Z'$. 
At least for this observable the narrow width method works well within the 
statistics; we have verified that this result also holds for other models. 

\vspace*{-0.5cm}
\nn
\vspace*{0.4mm}
\vspace*{-0.5cm}
\nn
\vspace*{0.4mm}

The rapidity ratio, $r_{y1}$, provides a complementary probe of the $Z'$ 
couplings. The relevant quantity to measure is the rapidity dependence of 
$Z'$ production cross section. Fig.~\ref{newdis} shows a comparison between 
the simplified ATLAS simulation and the narrow width approximation expectation 
which has been rescaled to go through the first Monte Carlo point. (Remember, 
we loose about 40$\%$ of our events due to the invariant mass cut on the lepton 
pair.) This 
result indicates that the narrow width approach does not do a very good job at 
getting the right shape for this distribution which results in values of 
$r_{y1}$ which are systematically high by as much as $30\%$ or more when this 
method is used. An examination of several other models with 
random $Z'$ masses and couplings shows similar qualitative results. 
Of course, we would need a more thorough simulation to 
verify these results and we would like to expand the study to many more models. 

One might ask how the narrow width approximation can do so well in the case of 
$A_{FB}$ but perform rather poorly for $r_{y1}$. It is clear that what is 
happening in the $A_{FB}$ case, since ratios of two cross sections at the 
{\it same} rapidity are taken, is that the excesses predicted by the narrow 
width method are cancelling out when the ratio is taken. Since the 
ratios at {\it different} rapidities 
are used in $r_{y1}$ this cancellation does not occur.

\vspace*{-0.5cm}
\nn
\vspace*{0.1mm}

In the narrow width approximation, assuming universality, $A_\tau$ provides a 
direct determination of the ratio of the left- and right-handed leptonic 
couplings of the $Z'$. In principle, this is a very sensitive probe of the $Z'$ 
couplings, \eg, in the ER5M as we vary the parameter $\theta$, $A_\tau$ takes 
on its entire allowed range of values and is generally large in magnitude. 
The technique is essentially that employed by LEP to extract this same 
quantity for the SM $Z$, however here we have to apply it in a hadronic 
environment. To study the $Z'\to \tau^+\tau^-$ requires good triggering for 
$\tau$-pairs with a excellent background rejection to get a clean sample. 
Studies by the CMS Collaboration indicate that these basic requirements can be 
achieved at a luminosity of $10^{33}$ at the LHC. 
A preliminary analysis of the use of $A_\tau$ to extract $Z'$ 
coupling information was performed some years ago by 
Anderson, Austern and Cahn{\cite {aac}} for the SSC. They concluded that a 
reasonable determination of $A_\tau$ might be possible for a $Z'$ with a mass 
near 1 TeV but that backgrounds became too serious if the mass were much 
larger. It would be very interesting and important to repeat this analysis 
for the LHC with a semi-realistic detector simulation to see if it remains 
valid. 

Other observables have been proposed to probe $Z'$ couplings:
\begin{itemize}
\item Associated $Z'$ production, \ie, $pp\to Z'V$ where $V=W,Z$ or $\gamma$. 
For different choices of $V$ different combinations of the $Z'$ couplings are 
being probed. The observable of interest here is the cross section ratio 
$R_V=\sigma (pp\to Z'V, Z'\to \ell^+\ell^-)/\sigma(pp\to Z'\to \ell^+\ell^-)$, 
wherein the $Z' \to \ell^+\ell^-$ branching fraction drops out. 
\item Rare $Z'$ decays such as the 3-body mode $Z'\to W\ell\nu_\ell$. 
The relevant observable 
here is the ratio of branching fractions, $r_{Wl\nu}$, for the $W\ell\nu_\ell$ 
final state scaled to the $\ell^+\ell^-$ discovery mode.
\item The ratio of cross sections for $pp\to Z'\to jj$ compared to 
$pp\to Z' \to \ell^+\ell^-$.
\end{itemize}
The immediate problem with the first two ideas is one of rate. For example, 
a 1 TeV $Z'$ in the ER5M has a value of $R_\gamma$ in the 0.001-0.007 range 
for photon $E_t$'s greater than 50 GeV with $|\eta_\gamma|\leq 2.5$. Both 
$R_Z$ and $R_W$ have similar magnitudes. Using leptonic $W,$ decay modes alone 
would compromise these measurements since the rates would be far too low. 
However,if the $Z,W\to jj$ modes are used we need to cleanly separate the 
two classes of events, thus requiring excellent hadronic mass resolution. 
While providing a reasonably clean signature, it does 
not seem too likely that associated production will be of much use for $Z'$ 
masses too far above 1 TeV due to a rapid fall off in statistics. A Monte 
Carlo analysis of these processes at the 
LHC needs to be performed.

The quantity $r_{Wl\nu}$ is generally found to be somewhat larger than $R_V$ 
for most EGM's and reasonable rates may be obtainable for $Z'$ masses as large 
as 1.5-2 TeV. The problem here is background since there is no  
$Z'\to \ell^+\ell^-$ in the final state to separate this from related SM 
processes. $S/B$ grows rapidly with increasing $Z'$ mass and it is unlikely 
that this mode can be used far above 1 TeV. Again, a Monte Carlo study of this 
and related processes at the LHC would be very useful. 

$Z'\to jj$ may be useful provided good resolution is available. The statistics 
is excellent but the QCD backgrounds are enormous. This possibility has 
already been explored in the somewhat tamer Tevatron environment by both the 
CDF and D0 
Collaborations{\cite {rharris,harris}} and has been briefly discussed by 
ATLAS{\cite {pog}}. It is clear that more detector studies need to be done to 
insure the usefulness of this mode.

\subsection{Lepton Colliders}

If a $Z'$ is sufficiently light that we can produce it directly at a lepton 
collider, the determination of its various properties will be straightforward. 
We need only to repeat the successful programs of SLC and LEP for the SM $Z$ 
over again at a higher energy. As noted above, however, it may be most likely 
that a $Z'$ will be too massive to undertake such a program at least at the 
first generation lepton colliders so that we can only make use of the same 
indirect signatures discussed above to sniff out the $Z'$ couplings. As we 
will see below, a major piece of the puzzle will be supplied if a hadron 
collider, such as the LHC, tells us the $Z'$ mass before coupling extraction 
analyses begin at lepton colliders.

\vspace*{-0.5cm}
\nn
\begin{figure}[htbp]
\leavevmode
\centerline{
\epsfig{file=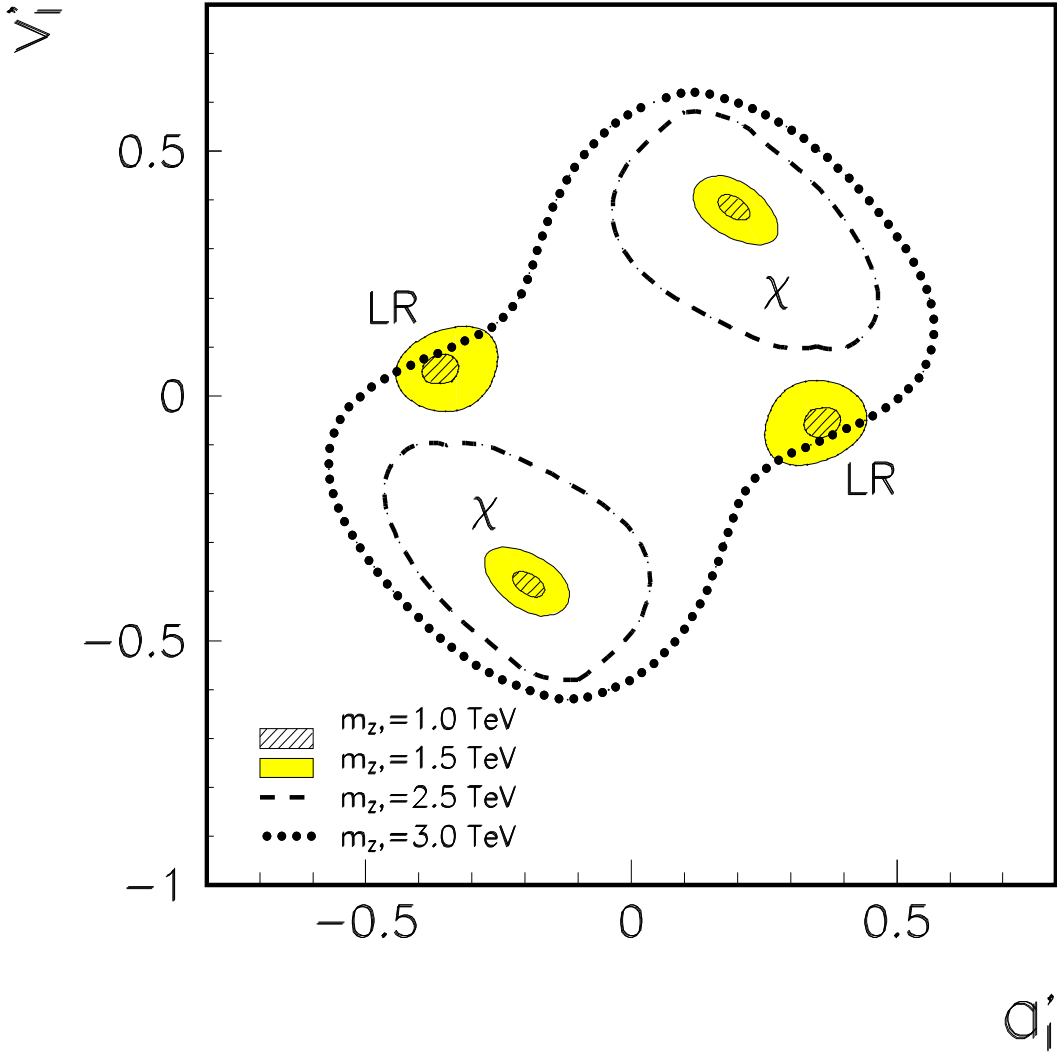,height=14cm,width=14cm,clip=}}
\vspace*{-0.3cm}
\caption{$95\%$ CL contours for $v'_l$ and $a'_l$ for a 500 GeV NLC with a 
luminosity of $50fb^{-1}$. The $Z'$ is taken to be in 
the $\chi$ or LRM with a 1(1.5) TeV mass corresponding to the hatched(shaded) 
area. The dashed(dotted) contours are $95\%$ CL limits on the $Z'll$ couplings 
for the $\chi$ case and a mass of 2.5(3) TeV. A beam polarization of 80$\%$ 
has been assumed.}
\label{sabine1}
\end{figure}
\vspace*{0.4mm}

In a contribution to these proceedings, Riemann{\cite {riemann}} analyzed the 
capability of future $e^+e^-$ colliders operating below the $Z'$ resonance 
to measure the $Z\bar ff$ couplings, where $f=\ell,b,c$. Her analysis 
implicitly assumed that the mass of the $Z'$ was already known and was used as 
an input into the numerical extraction of couplings. Fig.~\ref{sabine1} shows 
the capability of the NLC running at different energies to measure the 
leptonic couplings of the $Z'$ in the LRM and ER5M $\chi$ as the gauge boson 
mass is varied. It's clear from this analysis that with reasonable 
luminosities the NLC will be able to extract leptonic coupling information for 
$Z'$ masses up to $2-3\sqrt s$. (We recall that the {\it search reach} was 
found to be $6-10\sqrt s$.) 
These results are essentially statistics limited, 
there being few sizeable sources for systematic errors in the purely lepton 
mode. 

\vspace*{-0.5cm}
\nn
\begin{figure}[htbp]
\centerline{
\epsfig{file=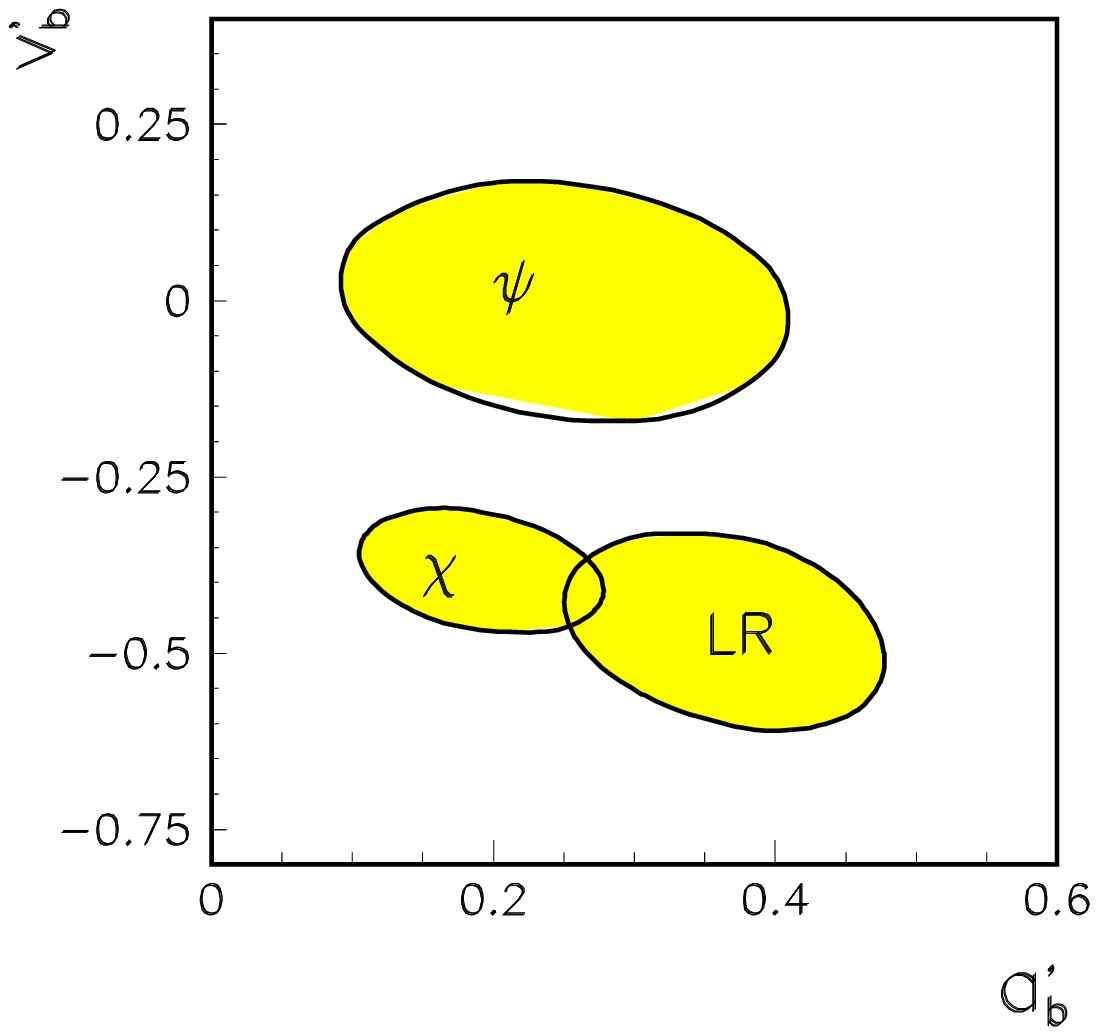,height=8.5cm,width=8.5cm}
\hspace*{-2mm}
\epsfig{file=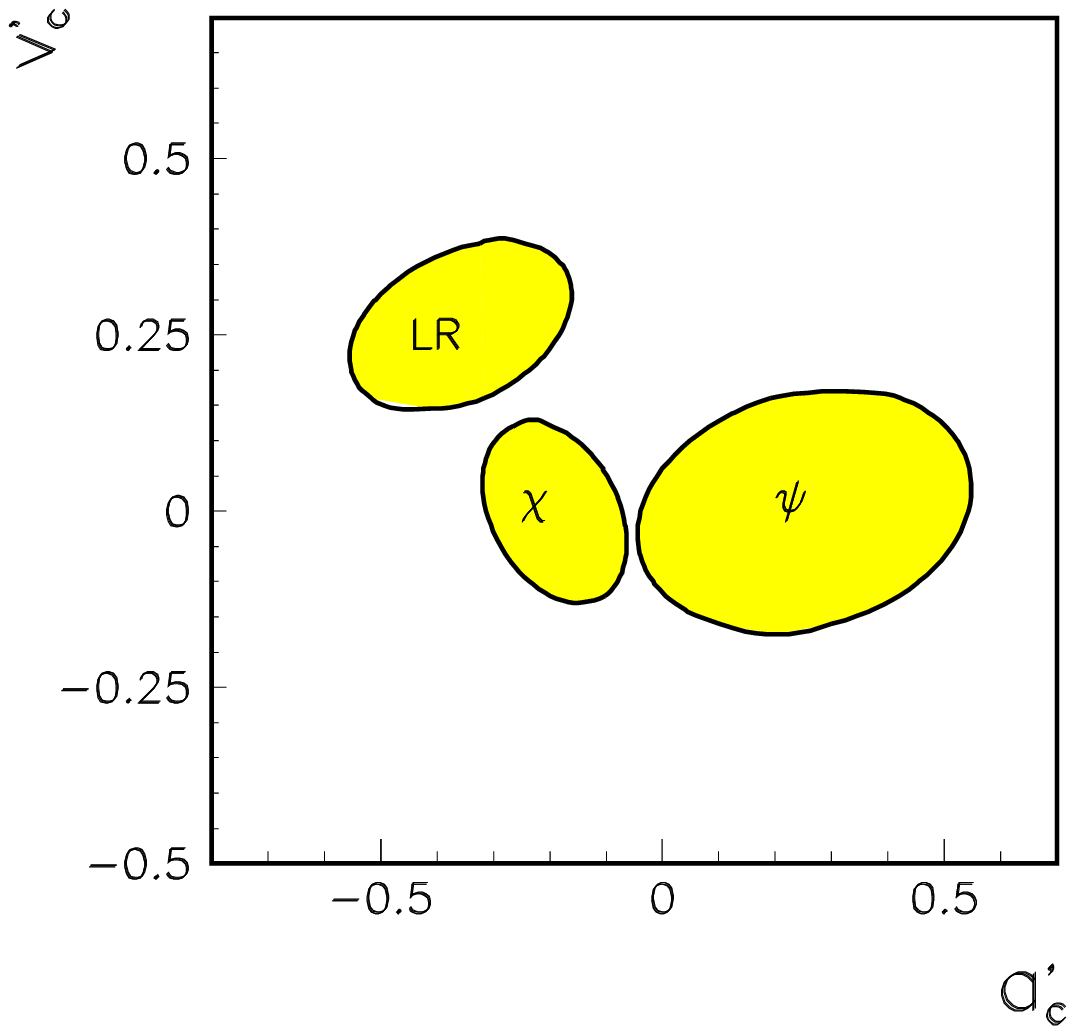,height=8.5cm,width=8.5cm}}
\vspace*{-0.2cm}
\caption{Model discrimination at $95\%$ CL, for a 1 TeV $Z'$ at a 500 GeV 
NLC with $50fb^{-1}$ of luminosity; on the left(right) for bottom(charm) 
quarks assuming a systematic uncertainty in observables of 1(1.5)$\%$. A 
$b(c)$-tagging efficiency of 60(40)$\%$ has been assumed together with a beam 
polarization of 80$\%$.}
\label{sabine2}
\end{figure}
\vspace*{0.1mm}

Riemann goes further in her analysis to take on the more daunting task of 
constraining the $c$ and $b$ quark couplings of the $Z'$. As she correctly 
points out, the size of the systematic errors for the measurements on these 
final states is rather critical to this program. For example, for a 
$Z_\chi$($Z_\psi$) with a 1 TeV mass, the size of the allowed region in the 
$v'_b-a'_b$($v'_c-a'_c$) plane approximately doubles at a 500 GeV NLC with a
luminosity of 50 $fb^{-1}$ if a systematic error of 1(1.5)$\%$ is added to all 
relevant observables. However, as Fig.~\ref{sabine2} shows, the NLC will 
still be 
able to extract coupling information and distinguish various models using the 
$c,b$ final states.

\vspace*{-0.5cm}
\nn
\begin{figure}[htbp]
\centerline{
\psfig{figure=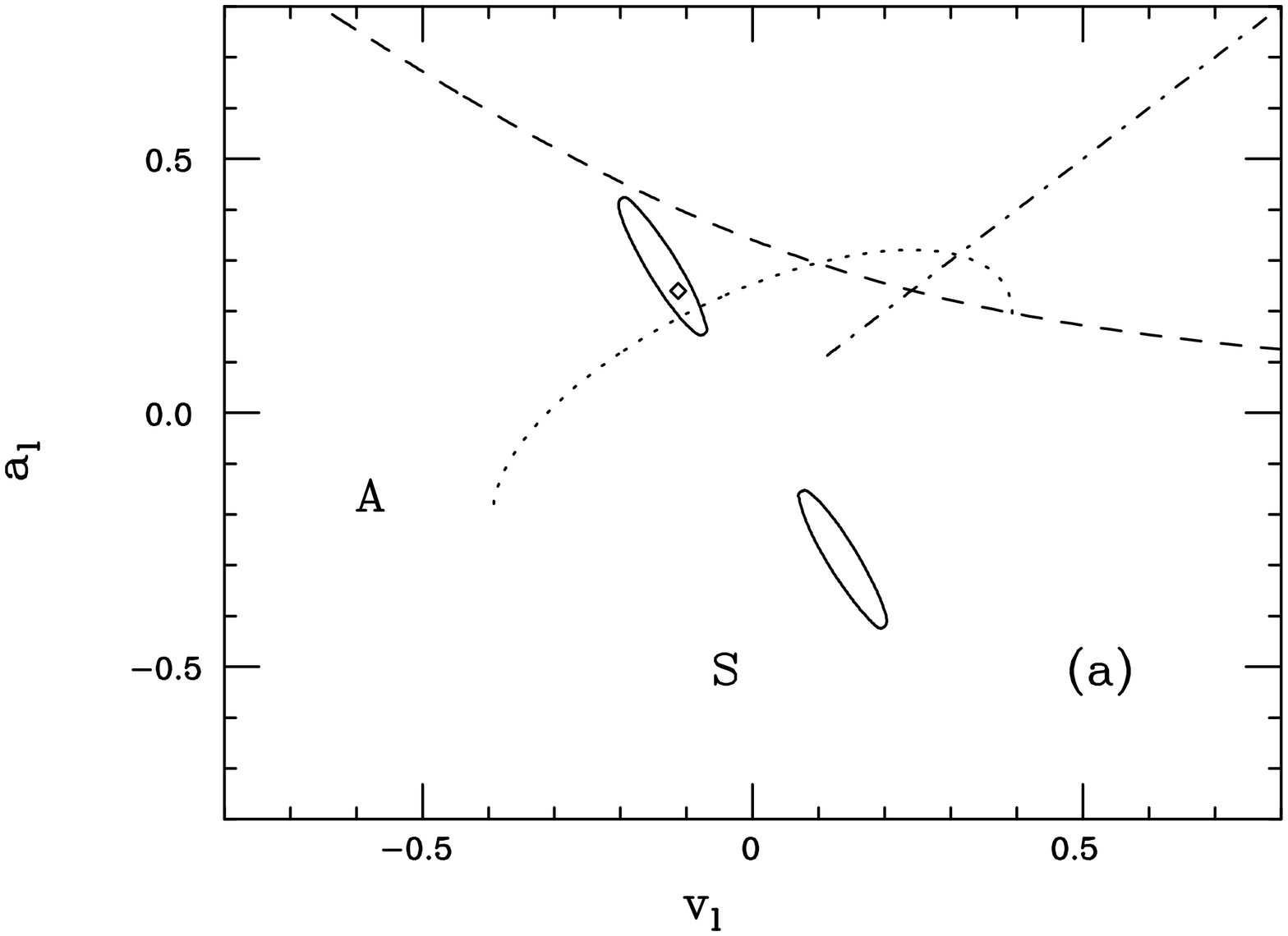,height=9.1cm,width=9.1cm,angle=-90}
\hspace*{-5mm}
\psfig{figure=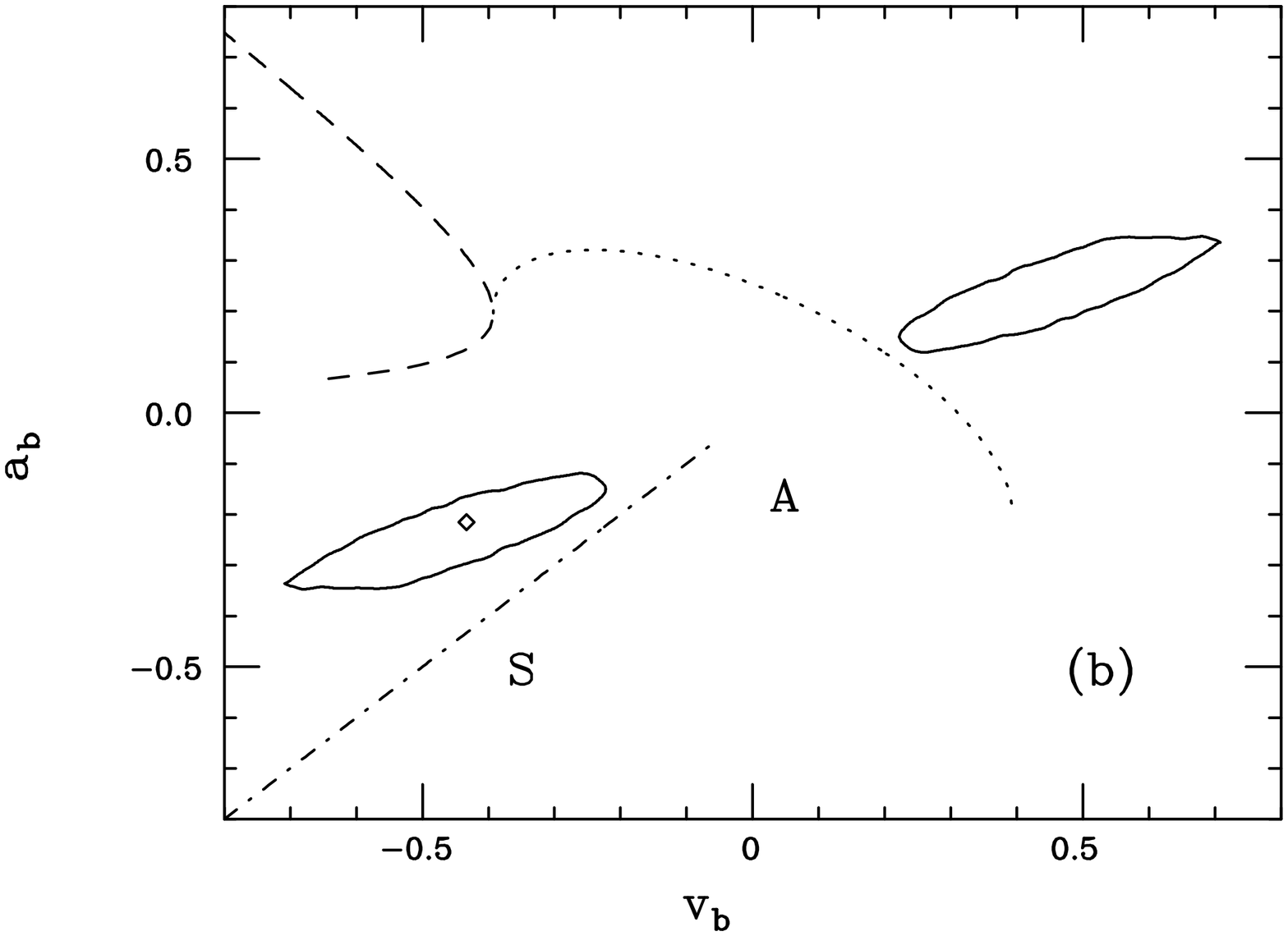,height=9.1cm,width=9.1cm,angle=-90}}
\vspace*{-0.75cm}
\centerline{
\psfig{figure=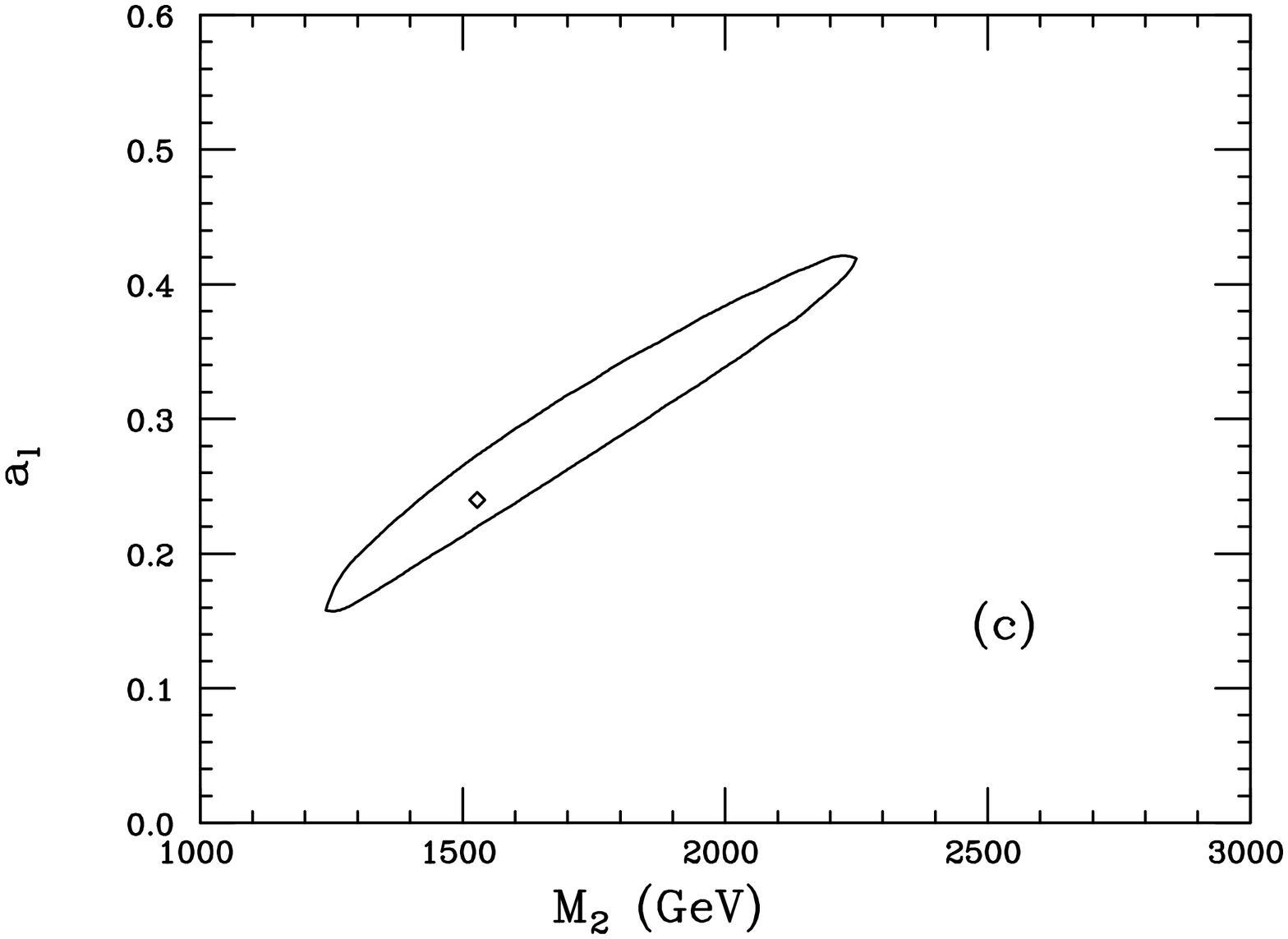,height=9.1cm,width=9.1cm,angle=-90}}
\vspace*{-0.3cm}
\caption{\small $95\%$ CL allowed regions for the extracted values of the 
(a) lepton and (b) $b$-quark couplings 
for a $Z'$ with randomly generated mass and couplings compared with the 
predictions 
of the $E_6$ model(dotted), the Left-Right Model(dashed), and the Un-unified 
Model(dash-dot), 
as well as the Sequential SM and Alternative LR Models(labeled by `S' and `A', 
respectively.) (c) Extracted $Z'$ mass; only the $a_\ell >0$ branch is shown. 
In all cases the diamond represents the corresponding input values. Here we 
seer that the couplings of this $Z'$ do not correspond to those of any of our 
favorite models.}
\label{tom1}
\end{figure}

What if the $Z'$ mass were not {\it a priori} known? It is clear in this 
circumstance that measurements taken at a single value of $\sqrt s$ will not 
be able to disentangle $Z'$ mass and coupling information. The reason is 
straightforward: to leading order in $s/M_{Z'}^2$, rescaling all of the 
couplings and the value of $Z'$ mass by a common factor would leave all of the 
observed deviations from the SM invariant. In this approximation, the $Z'$ 
exchange appears only as a contact interaction. Thus as long 
as $\sqrt s <M_{Z'}$, 
the only potential solution to this problem lies in obtaining data on the 
deviations from the SM predictions at several different values of $\sqrt s$ 
and combining them together in a single fit. In a presentation at this 
workshop, Rizzo{\cite {tom}} reported a first benchmark analysis of this kind 
in which data from different values of $\sqrt s$ are combined. Only the 
leptonic and $b$-quark couplings to the $Z'$ were considered. For $Z'$ 
masses in the 1.5-2 TeV range which were {\it a priori} unknown, this analysis 
found that combining data taken at 500, 750 and 1000 GeV was sufficient to 
determine the 4 unknown couplings as well as the $Z'$ mass. To insure 
model-independence, the mass and couplings were chosen {\it randomly} and 
{\it anonymously} from rather large ranges. 

\begin{figure}[htbp]
\centerline{
\psfig{figure=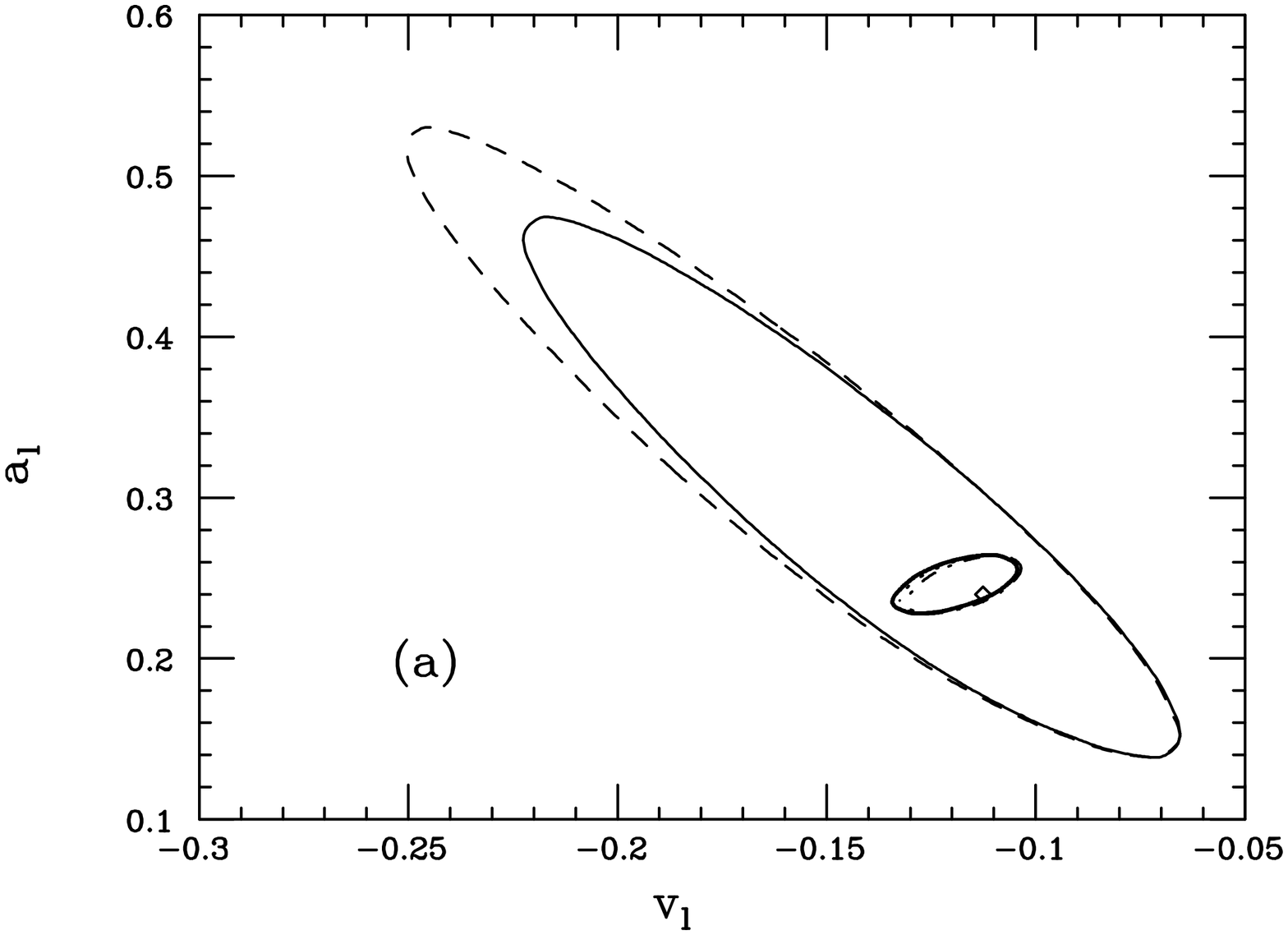,height=9.1cm,width=9.1cm,angle=-90}
\hspace*{-5mm}
\psfig{figure=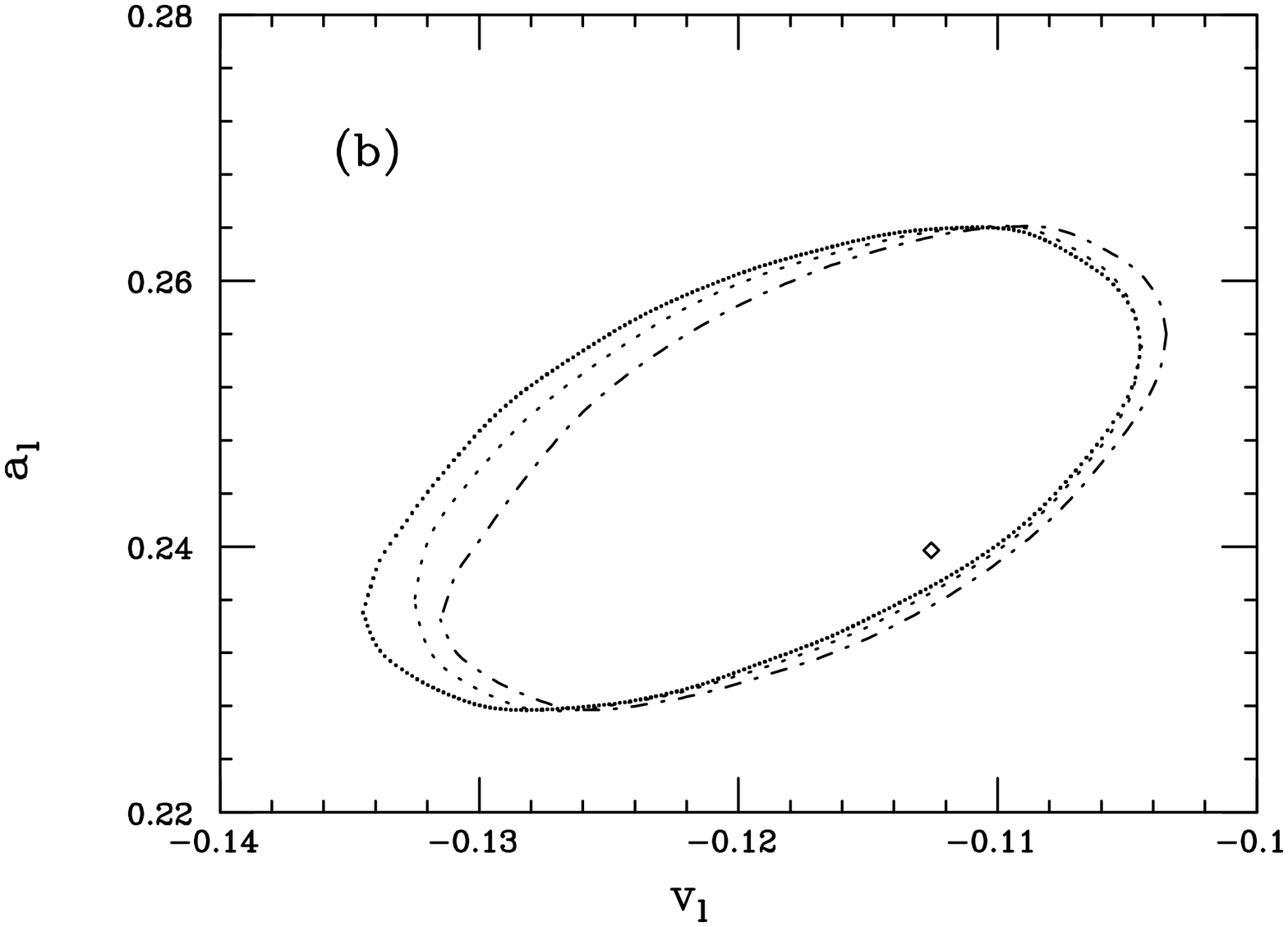,height=9.1cm,width=9.1cm,angle=-90}}
\vspace*{-0.75cm}
\centerline{
\psfig{figure=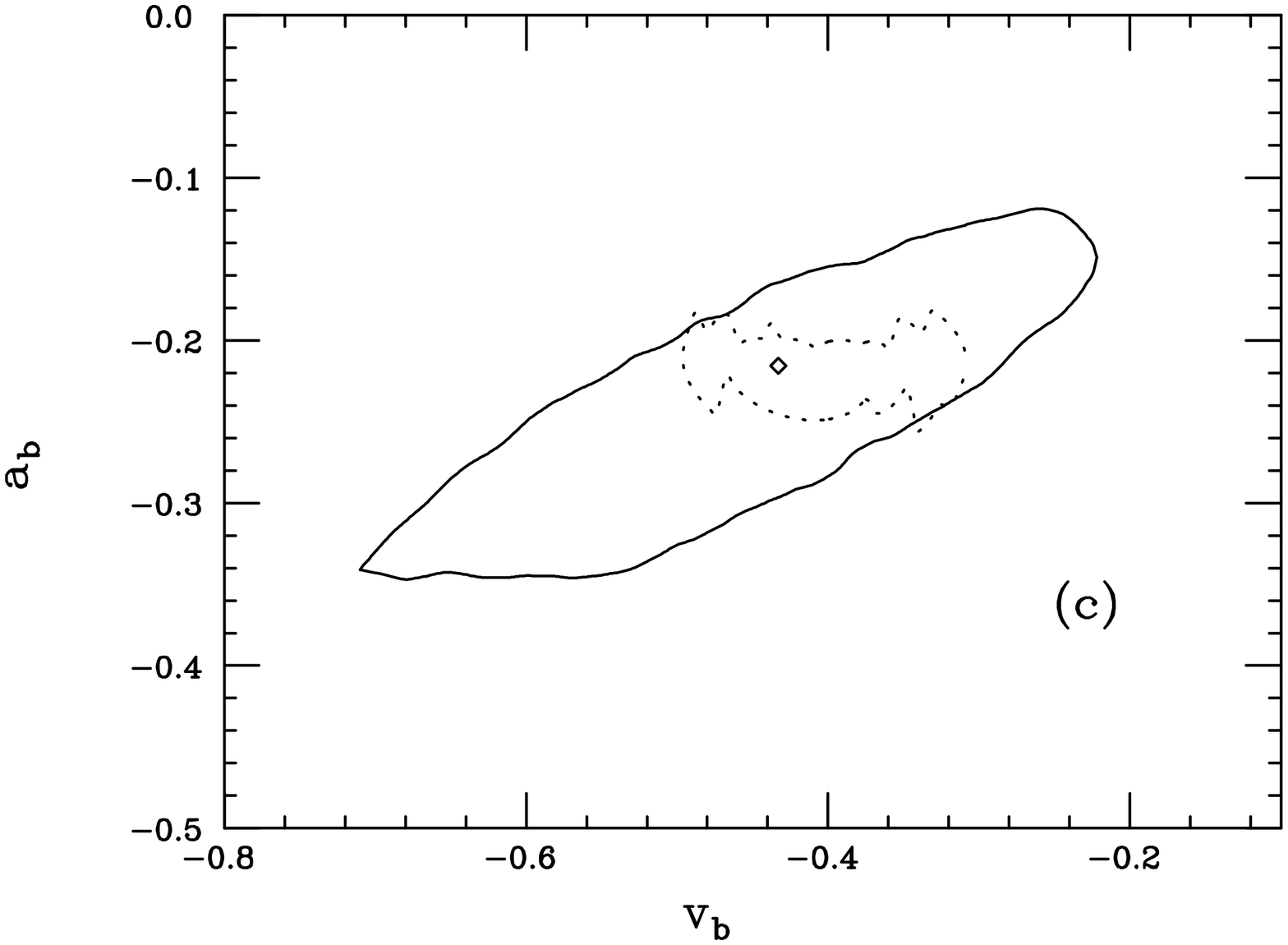,height=9.1cm,width=9.1cm,angle=-90}}
\vspace*{-0.7cm}
\caption{\small (a) Expanded lobe(solid) from the previous figure; the 
dashed curve shows 
the same result but for $P=80\%$. The smaller ovals, expanded in (b) apply 
when the $Z'$ mass is known. Here, in (b), $P=90(80)\%$ corresponds to the 
dash-dot(dotted) curve while the case of $P=90\%$ with $\delta P/P=5\%$ 
corresponds to the square-dotted curve. (c) Expanded lobe(solid) from the 
previous figure (b); 
the dotted curve corresponds to the case when $M_{Z'}$ is known.} 
\label{tom2}
\end{figure}

A sample result of this procedure is shown in Fig.~\ref{tom1}. The three 
figures correspond to two-dimensional projections of the full five dimensional 
($v'_l,a'_l,v'_b,a'_b,M_{Z'}$) $95\%$ CL fit. The following standard set of 
observables were employed: $\sigma_{f}$, $A_{FB}^{f}$, $A_{LR}^{f}$, 
$A_{pol}^{FB}(f)$ where $f=\ell,b$ labels the fermion in the final state and, 
special to the case of the tau, $<P_\tau>$ and $P_\tau^{FB}$. Universality 
amongst the generations was also assumed. While none of the couplings are 
extremely well determined we learn enough to rule out all conventional 
extended gauge models as the origin of this particular $Z'$. Note that 
knowledge of both 
the leptonic and $b-$ quarks couplings was required to rule out the case of 
an $E_6$ $Z'$. Fig.~\ref{tom2} shows how these results significantly improve 
if the the $Z'$ mass becomes known; one now performs a four dimensional fit 
instead of five. 

\vspace*{-0.5cm}
\nn

\subsection{$W'$ Couplings}

The model-independent extraction of the couplings of a new $W'$ have not 
attracted as much attention in the literature as has the $Z'$ case although 
several of the same techniques can be used. For example, the $A_{FB}$ of the
decay lepton from $W'$ decay can tell us a great deal about the $W'$ 
couplings. However, assuming these couplings are essentially chiral (as they 
are in all conventional models with a $W'$), as is well known this asymmetry 
will not be able to distinguish left-handed from right-handed couplings. As 
suggested by Cvetic and Godfrey{\cite {rev}}, the associated production of a 
$W'$ with a SM $W$ will only occur at a reasonable rate if the $W'$ has a 
substantial coupling to left-handed fields. For example, in the LRM, $WW'$ 
associated production cannot occur in the limit of zero gauge boson mixing if 
the quarks are assumed to be massless. In the Un-unified Model, however, the 
cross section for this process is rather large since the $W'$ couples in a 
left-handed manner. Similarly, rare decay modes such as $W'\to W f\bar f$ 
will not occur if the $W'$ is purely right-handed.  It would be quite 
beneficial if a model-independent analysis of the $W'$'s couplings could be 
performed.

Another way to get a handle on $W'$ couplings, particularly if the traditional 
lepton plus missing energy final state is suppressed, is to search for the 
decay $W'\to WZ$. This decay is particularly sensitive to the detailed 
structure of the extended gauge model. 
This analysis has already been done at the Tevatron by CDF 
for Run I{\cite {cdfw}} and has been extended for these proceedings by 
Toback{\cite {toback}} for the $W\to e\nu,Z\to jj$ decay mode. The $W'$ was 
assumed to have SM-like couplings to the initial $q\bar q$.

Apart from explicit factors which may appear at the $W'WZ$ vertex, the decay 
rate for $W'\to WZ$ scales as $M_{W'}^5$ in the large $W'$ mass limit. This is 
easily understood in that the $W'$ is actually coupling to the longitudinal 
components of the SM $W$ and $Z$ in this limit. Clearly, perturbation theory 
for the $W'$ width would become meaningless before the $W'$ mass exceeds 
values of order 1 TeV. A similar story applies to decays of the type 
$Z'\to WW$. Fortunately, in most realistic extended gauge models, the 
$W'WZ$ and $Z'WW$ vertex is only generated via $W-W'/Z-Z'$ mixing produced 
when we go over from the weak to mass eigenstate basis. In this case, the 
overall $W'WZ$[$Z'WW$] vertex is proportional to this mixing angle,  
which is generically of order $(M_W/M_{W'})^2$[$(M_Z/M_{Z'})^2$]. The growth 
in the $W'\to WZ$ and $Z' \to WW$ widths is thus significantly dampened and 
scales linearly with the mass of the new gauge boson. 

Assuming that the $W-W'$ mixing angle is just the ratio $(M_W/M_{W'})^2$, 
Toback shows that TeV33 has a significant sensitivity to this mode for $W'$ 
masses up to about 525 GeV for an integrated luminosity of $30 fb^{-1}$. A 
similar sensitivity was found for the $Z\to WW$ mode. It would be interesting 
to extend this study to the LHC.

\section{Summary/Outlook}

The physics of extended gauge sectors is particularly rich. Analyses have 
evolved to the point where detector considerations are becoming increasingly 
important. Many of the problems associated with the determination of the 
couplings of new gauge bosons now have to be faced with specific detector 
capabilities in mind. Although much work has been done, there is still a 
lot to be done along the directions begun here. Hopefully they will be 
completed before new gauge bosons are discovered.

\section{Acknowledgements}

The author would like to thank all of the members of this subgroup, as well as 
the entire New Phenomena group, for making the Snowmass meeting successful. 

\newpage
%
\def\MPL #1 #2 #3 {Mod.~Phys.~Lett.~{\bf#1},\ #2 (#3)}
\def\NPB #1 #2 #3 {Nucl.~Phys.~{\bf#1},\ #2 (#3)}
\def\PLB #1 #2 #3 {Phys.~Lett.~{\bf#1},\ #2 (#3)}
\def\PR #1 #2 #3 {Phys.~Rep.~{\bf#1},\ #2 (#3)}
\def\PRD #1 #2 #3 {Phys.~Rev.~{\bf#1},\ #2 (#3)}
\def\PRL #1 #2 #3 {Phys.~Rev.~Lett.~{\bf#1},\ #2 (#3)}
\def\RMP #1 #2 #3 {Rev.~Mod.~Phys.~{\bf#1},\ #2 (#3)}
\def\ZP #1 #2 #3 {Z.~Phys.~{\bf#1},\ #2 (#3)}
\def\IJMP #1 #2 #3 {Int.~J.~Mod.~Phys.~{\bf#1},\ #2 (#3)}


\begin{thebibliography}{99}

\bibitem{warsaw}
A. Blondel, plenary talk given at the {\it 28th 
International Conference on High Energy Physics}, Warsaw, Poland, 
25-31 July 1996. 

\bibitem{tev}
M. Pillai \etal, CDF Collaboration, hep-ex/9608006; 
S. Abachi \etal, D0 Collaboration, Fermilab report PUB-96/187-E. 

\bibitem{axi}
P. Frampton and S. Glashow, \PLB B190 157 1987 .

\bibitem{col}
R.S. Chivukula, A.G. Cohen, and E.H. Simmons, \PLB B380 92 1996 ;
E.H. Simmons, hep-ph/9608269 and these proceedings.

\bibitem{tc}
C. Hill, \PLB B345 483 1995 .

\bibitem{foot}
R. Foot and O. Hernandez, \PRD D41 946 1990 ;
R. Foot, O. Hernandez, and T.G. Rizzo, \PLB B246 183 1990 .

\bibitem{harris}
K. Cheung and R. Harris, these proceedings. 
 
\bibitem{rev}
For a complete set of references and a recent review of the physics of 
new electroweak gauge bosons see, 
M.\ Cvetic and S.\ Godfrey, Carleton University report OCIP/C-95-2, 1995,  
a part of the DPF long-range planning study to be published in
{\it Electroweak Symmetry Breaking and Physics Beyond the Standard Model}, 
eds. T.\ Barklow, S.\ Dawson, H.\ Haber, and J.\ Seigrist (World Scientific 
1996). See also, A.\ Djouadi, talk given at the {\it Workshop on Physics and 
Experiments with Linear Colliders}, Morioka-Appi, Japan, Sept. 8-12, 1995, 
hep-ph/9512311; J.\ Layssac \etal, hep-ph/9602327; A.\ Leike and S.\ Riemann, 
to appear in the {\it Proceedings of the Physics with $e^+e^-$ Linear 
Colliders Workshop}, Annecy-Gan Sasso-Hamburg, 1995, ed. P.\ Zerwas, 
hep-ph/9604321 and hep-ph/9607306.

\bibitem{physrep}
J.L. Hewett and T.G. Rizzo, \PR  183 193  1989 .

\bibitem{lykken}
J.D. Lykken, these proceedings. 

\bibitem{cl}
M. Cvetic and P. Langacker, \PRD D54 3570 1996 ~and \IJMP A11 1246 1996 .

\bibitem{farI}
A. Faraggi, \PLB B278 131 1992 .

\bibitem{farII}
A. Faraggi, \PLB B339 223 1994 .

\bibitem{fny}
A. Faraggi, D. Nanopoulos, and K. Yuan, \NPB B335 347 1990 .

\bibitem{chl}
S. Chaudhuri, G. Hockney, and J. Lykken, \NPB B469 357 1996 .

\bibitem{hl}
J. Lykken, ``String model building in the Age of D-branes'',
talk at the 4th International Conference on Supersymmetries
in Physics (SUSY 96), College Park,MD,29 May - 1 Jun 1996,
hep-th/9607144.

\bibitem{flip}
J. Lopez, D. Nanopoulos, and K. Yuan, \NPB B399 654 1993 ;
an earlier version is I. Antoniadis, J. Ellis, J. Hagelin,
and D. Nanopoulos, \PLB B231 65 1989 .

\bibitem{tgr}
T.G. Rizzo, these proceedings and hep-ph/9609248. 

\bibitem{toback}
D. Toback, these proceedings. 

\bibitem{wulz}
C.-E. Wulz, CMS Collaboration, these proceedings.

\bibitem{bsm}
T.G. Rizzo, \PRD D50 325 1994 ;
See also T.G. Rizzo in {\it Beyond the Standard Model IV}, Granlibakken CA, 
13-18 December 1994, p.24.

\bibitem{hr}
J.L. Hewett and T.G Rizzo in \IJMP A4 4551 1989  ~and  in {\it Proceedings of 
the DPF Summer Study on High Energy Physics in the 1990's}, Snowmass, CO, 
July 1988, ed. by S. Jensen(World Scientific, Singapore 1989), p.235. 

\bibitem{lmc}
See, {\it $\mu^+\mu^-$ Collider Feasibility Study}, BNL report BNL-52503, 1996. 

\bibitem{tatc}
E. Eichten and K. Lane, Fermilab-Conf-96/298-T, BUHEP-96-34, hep-ph/9609298 
and these proceedings; C. Hill and S. Parke, \PRD D49 4454 1994  ;
G. Burdman, these proceedings; C. Hill, \PLB B345 483 1995 .

\bibitem{cat}
S. Chivukula and J. Terning, \PLB B385 209 1996 .

\bibitem{godfrey}
S. Godfrey, these proceedings.

\bibitem{fc}
A. Leike, \ZP C62 265 1994 ;
D. Choudhury, F. Cuypers and A. Leike, \PLB B325 500 1994 ;
F. Cuypers, hep-ph/9602426.

\bibitem{cuy}
F. Cuypers, these proceedings. 

\bibitem{wrref1}
See, for example, J. Maalampi, A. Pietila and J. Vuori, \PLB B297 327 1992 ;
T.G. Rizzo, \PRD D38 71 1988 .

\bibitem{wrref2}
T.G. Rizzo, \PLB B192 125 1987 ~and SLAC-PUB-6591(hep-ph/9407367), 1994. 

\bibitem{wrref3}
T.G. Rizzo, \PRD D50 5602 1994 .

\bibitem{wrref4}
K. Huitu, J. Maalampi and M. Raidal, \PLB B365 407 1996 .

\bibitem{joa}
J.L. Hewett, these proceedings.

\bibitem{joa2}
J.L. Hewett, work in progress.

\bibitem{old}
For original references, see the review by Cvetic and Godfrey in 
Ref.{\cite {rev}}.

\bibitem{aac}
J.D. Anderson, M.H. Austern and R.N. Cahn, \PRD D46 290 1992 .

\bibitem{rharris}
F. Abe, \etal, CDF Collaboration, \PRL 71 2542 1993 ~and \PRL 74 3538 1995 ;
F. Abe, \etal, CDF Collaboration, FERMILAB-PUB-96-317-E, 1996; 
S. Abachi, \etal, D0 Collaboration, FERMILAB-CONF-96-168-E, 1996.

\bibitem{pog}
A. Henriques and L. Poggioli, ATLAS Collaboration, Note PHYS-NO-010, 1992; 
see also, T.G. Rizzo, \PRD D48 4236 1993 .

\bibitem{riemann}
S. Riemann, these proceedings.

\bibitem{tom}
T.G. Rizzo, SLAC-PUB-7151 and SLAC-PUB-7250, 1996.

\bibitem{cdfw}
T. Kamon, CDF Collaboration, {\it Proceedings of the XXXIst Recontres de 
Moriond(QCD)}, Les Arces, Savoie, France, March 23-30 1996, 
Fermilab-Conf-96/106-E.

%
\end{thebibliography}
\end{document}